\documentclass[useAMS,usenatbib]{mn2e}

\usepackage{epsfig}
\usepackage{myaasmacros}
\usepackage{amssymb}
\title[Surface density profiles of collisionless disc merger remnants]
{Surface density profiles of collisionless disc merger remnants}
\author[Thorsten Naab \& Ignacio Trujillo]{Thorsten Naab$^{1,2}$\thanks{E-mail:
naab@usm.lmu.de} and Ignacio Trujillo$^{3,4}$
\\
$^{1}$Insitute of Astronomy, Madingley Road, Cambridge, CB3 0HA, UK\\
$^{2}$ Universit\"ats-Sternwarte M\"unchen, Scheinerstr.\ 1, D-81679 M\"unc
hen,Germany;\\
$^{3}$Max-Planck-Institut f\"ur Astronomie, K\"onigstuhl 17,
Heidelberg D-69117, Germany\\
$^{4}$School of Physics and Astronomy, University of Nottingham, University
Park, Nottingham NG7 2RD, UK}

\begin{document}

\date{Accepted ???. Received ??? in original form ???}

\pagerange{\pageref{firstpage}--\pageref{lastpage}} \pubyear{2002}

\maketitle

\label{firstpage}

\begin{abstract}
We present a detailed surface density analysis of a large sample of simulated
collisionless mergers of disc galaxies  with bulges (mass ratios 1:1, 2:1, 3:1,
4:1, and 6:1) and without bulges  (mass ratios 1:1 and 3:1). A dissipative component 
was not included. The randomly projected remnants were fit with a single  
S\'ersic function and a S\'ersic
function plus an exponential. They were classified, according to their
bulge--to--total ($B/T$) ratio, either as a one-component system or
as a two-component systems.  In general projection effects change
the classification of a remnant. Only merger remnants  of discs with bulges
show properties similar to observed early--type galaxies.  Their B/T ratios are
in the range $0.2<B/T<1$.   Surprisingly, the initial mass ratio has a weak
influence on the distributions  of $B/T$, effective radius and S\'ersic index
$n$. For all one-component projections ($\approx 60\%$ of all
projections)  the S\'ersic index distribution peaks at $3<n<4$. However, the mass ratio 
is tightly linked  to the properties of the outer exponential
components which resemble pressure supported,  spheroidal halos for 1:1 and 2:1
remnants and elongated heated discs for 6:1 remnants.  We found distinct
correlations between the fitting parameters which are very similar to  observed
relations (e.g. larger bulges have lower  effective surface densities).
No indications for a correlation between the surface density profiles
and other global parameters  like remnant masses, isophotal shapes or central
velocity dispersions are found.  The remnants have properties similar to
giant elliptical galaxies in the  intermediate mass regime. A binary disc merger origin for
all early-type galaxies, especially the most massive ones, is unlikely.
Observed nearby merger remnants have properties similar to the simulated
remnants. They can have formed from binary disc mergers and might evolve into
early--type galaxies within  a few Gyrs.
\end{abstract}


\section{Introduction}
\label{INTRODUCTION}

For many decades the surface brightness profiles of elliptical  galaxies have been considered to be
well described by the \citet{1948AnAp...11..247D} r$^{1/4}$-profile. However,  from the early
1990's, as the quality of  photometric data improved, it became clear that the de Vaucouleurs law
is not valid over  the whole observable radius range and it became necessary to use a  fitting
function which can account for variations  in the curvature of the light profile 
\citep{1987IAUS..127...47C, 1988MNRAS.232..239D,1993MNRAS.265.1013C,  1993A&A...278...23B}.
Following these pioneering works, several authors have  showed that the surface brightness profiles
of  elliptical galaxies are better described by the \citet{1968adga.book.....S} r$^{1/n}$-profile.
This is not surprising as the shape parameter $n$ of the S\'ersic function is an additional
fitting parameter.  Interestingly, $n$ is  strongly correlated with other
observed global properties of elliptical galaxies derived independently of the r$^{1/n}$ fits, such
as: total luminosity, effective radius,   \citep{1993MNRAS.265.1013C,1994MNRAS.268L..11Y,
1995MNRAS.273.1141Y, 1997neg..conf..239J,1997A&A...321..111P}, central velocity dispersion
\citep{2001AJ....122.1707G,2004ApJ...601L..33V}, and also the masses of super-massive black holes
\citep{2001ApJ...563L..11G}. The non--homology of elliptical galaxies has also been shown to account 
for a large fraction of the tilt of the Fundamental Plane
\citep{2004ApJ...600L..39T}. In general, more massive giant ellipticals have more  concentrated
light profiles indicated by larger values of $n$ which is most likely a  direct consequence of the
formation process. Recently, \citet{2004MNRAS.355.1155D}  have performed a two component
decomposition (inner S\'ersic and outer exponential "disc")  of a large sample of 558 early--type
galaxies. They found strong evidence for outer  exponential components in early--type galaxies at
all luminosities.  However, as they have pointed out, it was not possible to prove without further 
information about kinematics and stellar populations whether this analysis has any physical
significance. In particular it was not clear whether the exponential component  represented a
classical flattened, rotationally supported thin or thick disc. If the exponential components turn out 
to be real  \citet{2004MNRAS.355.1155D} found possibly meaningful correlations  between
the fitting parameters such as: bulge sizes and disc sizes; bulge-to-disc ratio,  bulge shape
parameter, and bulge effective radius. 

It is the aim of this paper to test the significance of binary disc mergers  for the formation and
evolution of early--type galaxies and bulges by investigating  the surface density profiles of the
merger remnants. As our progenitor discs have  properties similar to present day disc galaxies a
simulated merger remnant will not resemble a present day early--type in the sense that its 
stellar population is too young. The distribution and kinematics of the stars, however, 
might be in agreement. If the discs have merged early enough, $ z \gtrsim 1$, 
having similar properties (and there are indications for the existence of large scale discs 
at high redshift, see e.g. \citealp{2003ApJ...591L..95L}) their stellar populations 
might have evolved into early--type populations by now (discs might on average have been  smaller
in the past but our models are scale-free) and a comparison to the properties of  present 
day early--type galaxies, as mentioned above, is justified. On the other hand  with a direct 
comparison to observed present day merger remnants we can test whether nearby 
merger remnants originate from discs and whether they will evolve into early--type 
galaxies in the future. 

Such a sample has been published by \citet{2004AJ....128.2098R} who have successfully  fitted
S\'ersic-profiles to surface brightness profiles of 51 observed merger remnants.  They conclude
that the remnants might evolve into ellipticals as they can have (K-band)  luminosities in the
range of an $L^*$ elliptical and photometric properties  that are in principal similar to observed
ellipticals. Most of the remnants have  S\'ersic indices between $ 1 < n  < 6$ with a peak at
$n=2$. They find  indications for more concentrated profiles, $n > 8$ for some remnants.  However,
there are still some open questions, e.g. they do not find a  correlation of the shape parameter
with luminosity or isophotal shape (most of  the remnants have discy isophotes) and a kinematical 
investigation is still missing and at present it seems still difficult to argue that  present disc
merger remnants evolve into bona fide early--type galaxies based on observations alone.

On the theoretical side it is now almost 30 years that it has been proposed that elliptical
galaxies might form from mergers of disc galaxies. As an alternative
to the even older monolithic collapse scenario the 'merger hypothesis'
became more and more popular in the framework of cosmological models
where structure forms hierarchically by mergers and every galaxy is 
expected to have experienced at least one (maybe more) major merger during 
its lifetime. The first numerical experiments in the early 1970s have 
been followed by more and more self-consistent collisionless major merger 
models of disc galaxies (see \citealp{2003ApJ...597..893N} and references therein). 
As pure disc galaxies have much lower central phase-space densities 
than ellipticals it has been argued for a long time that
one can not make elliptical galaxies by collisionless mergers of pure discs  
\citep{1980ComAp...8..177O,1986ApJ...310..593C}.
However, \citet{1993ApJ...409..548H} ``solved'' this problem by adding 
sufficiently massive concentrated bulge components to the progenitor galaxies. 
The surface density profiles of the merger remnants have been compared 
to a de Vaucouleurs $r^{1/4}$-profile and a reasonable agreement has been found
\citep{1992ApJ...393..484B,1992ApJ...400..460H,1993ApJ...409..548H}. 
\citet{2003ApJ...597..893N} have found a good statistical 
agreement between observations of intermediate mass giant ellipticals 
and simulations mergers of discs with varying mass-ratios with respect to
 kinematic and photometric properties including the isophotal shape. The most 
characteristic change with mass ratio is the amount of rotation of the remnants. 
Equal mass remnants are slowly rotating whereas unequal mass remnants are in general fast rotators 
\citep{1998giis.conf..275B,1999ApJ...523L.133N,2003ApJ...597..893N}. There are, however, 
disagreements in the higher order moments of the line-of-sight velocity 
distributions for fast rotating remnants \citep{2001ApJ...555L..91N}. Orbital analysis 
of merger remnants has shown that equal mass remnants are dominated by box orbits 
whereas unequal mass remnants are dominated by tube orbits 
(see \citealp{2005MNRAS.360.1185J}, and references therein). As the stellar orbits are
the backbones of elliptical galaxies a different orbital content might lead to different 
projected surface density profiles. 

A statistical analysis of the surface density 
profiles of merger remnants has not been performed so far. Given the observational 
evidence that the S\'ersic-function provides a better fit 
to observed ellipticals, we decided to investigate the surface 
density profiles of collisionless disc merger remnants in more detail 
and to probe whether the profiles of the simulated galaxies are well described 
by a S\'ersic function or if they can be decomposed in a S\'ersic part and a part 
with an exponential profile (here called "disc" for convenience)  and 
whether the observed relations between the structural 
parameters can be recovered. Only recently, some groups have used the 
S\'ersic-function to fit simulated merger remnants. \citet{2005MNRAS.357..753G}
found that merger remnants of disc galaxies with massive bulges are more 
concentrated than pure disc mergers. \citet{2004A&A...418L..27B} realized that 
 the outer parts of remnants with mass ratios larger than 4:1 can show
an exponential surface density profile similar to a disc galaxy. 

In this paper we present the detailed application of
 an observational analysis method \citep{2001AJ....122...38T,2002MNRAS.333..633A} 
for surface brightness profiles of 
early--type galaxies to the projected surface density profiles of 
simulated galaxies. The profiles are fitted with a pure S\'ersic and a 
S\'ersic + exponential profile taking the full photometric information 
into account (e.g. seeing, ellipticities, and errors). 
This enables us to perform a self-consistent disc + bulge decomposition. 
In a first step, the analysis  is applied to a statistical set of simulated 
collisionless merger remnants of progenitor disc galaxies with bulges 
and mass ratios of 1:1, 2:1, 3:1, 4:1, and 6:1 and 
bulge-less progenitors with mass ratios of 1:1 and 3:1, for comparison. 

The paper is structured as follows: In Section \ref{PARAMETERS} we review the 
parameters of the simulations used.  The surface density analysis is described
in detail in Section \ref{CLASSIFICATION} with an exemplary application 
to the initial conditions and individual simulated merger remnants  shown 
in the appendix. In Section \ref{STATISTIC} we present the results 
based on the whole sample. The resulting parameter relations are discussed 
separately in Section \ref{GLOBAL}. In Section \ref{NATURE} we investigate 
the nature of the outer exponential component and in Section \ref{CONCLUSIONS} 
we summarise and discuss the results.  
\section{Simulation parameters} 
\label{PARAMETERS} 
The statistical set of merger simulations of disc galaxies with mass
ratios of 1:1, 2:1, 3:1 and 4:1  is described in detail in
\citet{2003ApJ...597..893N}. Here we only give a brief summary.
We used the following system of units: gravitational constant G=1,
exponential scale length of the larger disc $h=1$ and mass of the
larger disc $M_d=1$. Each galaxy consists of an exponential disc, a
spherical, non-rotating bulge with mass $M_b = 1/3$, a Hernquist
density profile \citep{1990ApJ...356..359H} with a scale length
$a_b=0.2h$  and a spherical pseudo-isothermal halo with a mass
$M_h=5.8$, cut-off radius $r_c=10h$ and core radius $\gamma=1h$
\citep{1993ApJS...86..389H}. The equal-mass mergers were calculated
adopting in total 400000 particles with each galaxy consisting of
20000 bulge particles, 60000 disc particles, and 120000 halo
particles. The low-mass companion contained a fraction of $1 / \eta $
(mass ratio $\eta$ = 1/1, 2/1, 3/1, 4/1 and 6/1) the mass and the number
of particles in each component with a disc scale length of
$h=\sqrt{1/\eta }$, as expected from the Tully-Fisher relation
\citep{1992ApJ...387...47P}. The galaxies approached each other on
nearly parabolic orbits (in agreement with estimates from cosmological simulations 
\citep{2006A&A...445..403K})
 with an initial separation of $r_{sep} = 30$
length units and a pericenter distance of $r_p = 2$ length units. The
initial parameters for the disc inclinations were selected in an
unbiased way following the procedure described by
\citet{1998giis.conf..275B}. We have realized 16 different initial
orientations resulting in 16 equal-mass remnants and 32 remnants for
every higher mass ratio, respectively (see Tbl. 1 in \citealp{2003ApJ...597..893N}). 
In addition, we have re-run
the merger simulations with mass ratios of 1:1 and 3:1 with bulge-less
progenitor discs. This resulted in 192 simulated merger remnants that are 
investigated in this paper.    

All simulations have been performed using the tree-code VINE 
(Wetzstein et al., in preparation) in combination
with the GRAPE-5/6 special purpose hardware
\citep{2000PASJ...52..659K}. The Plummer-softening for the force
calculation was set to $\epsilon = 0.05$. Time integration was
performed  with a with a Leap-Frog integrator with a fixed time-step
of $dt = 0.04$. For further details on the simulations see 
\citet{2003ApJ...597..893N}.

\section{Analysis of the surface density profile and classification}
\label{CLASSIFICATION}
In the classical view the surface brightness profiles of elliptical
galaxies have been assumed to follow the de Vaucouleurs $r^{1/4}$-profile 
\citep{1948AnAp...11..247D} 
\begin{equation} 
I(r)=I(0)e^{-7.67(r/r_e)^{1/4}}.
\end{equation}
$I(0)$ is the central surface brightness, $r_e$ is semi-major axis effective radius
and the factor $7.67$ is chosen such that the effective radius
encloses half the total luminosity of the galaxy. In this paper we use the 
S\'ersic-function \citep{1968adga.book.....S} 
to fit the surface density profile  
of simulated merger remnants. The S\'ersic-function is a generalized
version of the de Vaucouleurs-profile allowing the exponent to account 
for the varying curvature of the surface brightness profiles of observed galaxies. 
As we have shown in Sec. \ref{INTRODUCTION} there is strong evidence  that the curvature 
of the light profile is a physically meaningful parameter which is closely linked 
to other properties of ellipticals. 
Throughout the paper we assume that the surface density
of the simulated merger remnants reflect their surface brightness
with a constant mass--to--light ratio along the radial 
profile. The S\'ersic-function can be written as   
\begin{equation} 
I(r)=I(0)e^{-b_n(r/r_e)^{1/n}}. \label{ser_fit}
\end{equation}
The three free parameters (which is one more than for the de Vaucouleurs-profile) 
are: the central surface brightness $I(0)$, the effective radius
$r_e$, and the shape parameter $n$, the so called S\'ersic-index. The
factor $b_n$ is a function of the shape parameter $n$, and is chosen
such that the effective radius encloses half of the total luminosity. The
exact value is derived from $\Gamma$(2n)=2$\gamma$(2n,b$_n$), where
$\Gamma$(a) and $\gamma$(a,x) are the gamma and the incomplete gamma
functions. In case of $n=1$ the profile is exponential and in case of
$n=4$ the profile is de Vaucouleurs.     

To systematically analyse the simulated merger remnants we created 
artificial images of every remnant seen from 100 random projections by 
binning the central 10 length units into $128 \times 128$
pixels. For the simulation parameters given above this resulted in
14400 images for mergers with bulges and 4800 for mergers without bulges
in the progenitor galaxies. In an automated procedure every image was
smoothed with a Gaussian filter of standard deviation 0.05 length scales. 
The surface density profile, the isophotes and their best fitting ellipses
 were determined using a data reduction package kindly provided by Ralf
Bender. The surface density profile was thereafter fitted as described
below using the algorithm presented in \citet{2001AJ....122...38T} and
\citet{2002MNRAS.333..633A}. As the S\'ersic-function fit is sensitive to
the detailed shape in the innermost regions of the surface density
profiles we followed a conservative approach and excluded all data
points at radii $r_{fit} < 0.1$ ($\approx 2$ softening lengths) which 
might be affected by the numerical force softening. However, using 
slightly smaller or larger values for $r_{fit}$ did not change 
the global results.

In addition to the fit of a single S\'ersic-function to the surface density profiles we performed a
decomposition (usually called "bulge-disc decomposition")  allowing for a bulge component described
by a S\'ersic-function and an exponential component like in  disc galaxies \citep{1970ApJ...160..811F}: 
\begin{eqnarray}  
I(r) & = & I_{Disc}(r)+I_{Bulge}(r) \nonumber \\ & = & I_{D}(0)\exp{(r/h_D)} +
I_{B}(0)e^{-b_n(r/r_{eB})^{1/n_B}} \label{DB_fit}. 
\end{eqnarray}  
The meaning of the structural
parameters of the bulge component are the same as those described for the S\'ersic only model.
$I_{D}(0)$ is the central surface  density and $h_D$ is the scale length of the exponential
component.  The set of free parameters is completed with the ellipticities of the bulge
$\epsilon_B$ and the disc $\epsilon_D$. To avoid confusion with the observational notation 
all surface densities that result directly from the simulated remnants are denoted by $\Sigma$ in the 
following. We have to note at this point that an exponential profile
alone does not prove the  existence of a disc component which in addition has to be flattened and
supported by  rotation. In Section \ref{NATURE} we will show that some remnants have exponential
components  that are not discs. For convenience we keep the notation "disc" for the  exponential
component as long as there is not other evidence than the profile shape.

To quantify which fraction of mass corresponds to the bulge and which
fraction corresponds to the disc we used the  bulge--to--total ($B/T$)
ratio. $B$ is the projected bulge mass and $T=B+D$ is the total mass
(here $D$ is the disc mass) of the system. Using the parameter values
obtained from the fit we estimated $B/T$ according to:
\begin{eqnarray}
\frac{B}{T}=\frac{I_{B}(0)r_{eB}^2(n_B/b_n^{2n_B})(1-\epsilon_B)
\Gamma(2n_B)}{I_{B}(0)r_{eB}^2(n_B/b_n^{2n_B}) \Gamma(2n_B)(1-\epsilon_B)
+I_D(0)h_D^2(1-\epsilon_D)}
\label{BT_ana}
\end{eqnarray}
where $\Gamma(a)$ is the gamma function. The frequently used
bulge--to--disc ratio, $B/D$, is then given by  
\begin{eqnarray}
B/D=\frac{B/T}{1-B/T}. 
\end{eqnarray}

In some rare cases (below 2\%) the application of Eqn. \ref{BT_ana} directly using the
parameters of the fit caused problems. For example, if the fitted central 
surface density of the exponential scale component is below the outermost 
measured value of the remnant and the scale length $h_D$ becomes 
unrealistically large (i.e. larger than the size of the remnant 
containing 95\% of all the particles), we would significantly overestimate 
the disc mass and underestimate $B/T$.  For these cases we assumed $B/T=1$. 

The size of the $BD$ systems was estimated by computing the global
semi-major effective radius, $r_{eG}$, solving the equation
\begin{equation}
\frac{B/T}{1-B/T}
\biggr[1-\frac{2\gamma(2n,b_n(r_{eG}/r_{eB})^{1/n})}{\Gamma(2n)}\biggr]=
2\gamma(2,r_{eG}/h_{D})-1. \label{re_global}
\end{equation}
A $BD$--decomposition of the merger remnants allowed us to
classify the surface density profiles in two categories depending on 
the bulge--to--total ($B/T$) ratios. Projections with $B/T > 0.7$ were
considered pure bulge systems and, for simplicity reasons, the
structural parameters determined from the S\'ersic-only fits were
used. Projections with $B/T < 0.7$ were considered two-component
disc+bulge systems. For all cases we compared the reduced chi--squared
of the S\'ersic--only fit and the $BD$--decomposition fit. In general the fit 
was considered sufficiently accurate if the reduced chi--squared  
value of the fit was smaller than unity. For those cases where both fits resulted in 
a reduced chi--squared  value smaller than unity (see for example second row in Fig. 
\ref{61_surf_nacho}) we computed the probability 
\begin{equation}
P(\chi^2|\nu) = \gamma \left(\frac{\nu}{2},\frac{\chi^2}{2}\right)
\label{prob}
\end{equation}
that the chi--squared value is due to chance \citep{1992nrfa.book.....P}. 
Here $\gamma$ is the incomplete gamma function 
and $\nu$ is the number of degrees of freedom. If that probability for the S\'ersic-only fit
was smaller than $P < 0.32$ the fit was considered as sufficiently accurate and the 
projected remnant was classified as one component with $B/T =1$. Changing the threshold probability to 
$P=0.5$ or $P=0.2$ does not change the results.  

As it is the aim of this study to investigate the statistical properties of the 
remnants we apply identical boundary conditions like weights, computation 
of errors etc. to all projected images even if fine-tuning might reduce 
the residuals slightly in one or the other case.

A potential source of concern for fitting a bulge and a disc to the surface
brightness distribution of the galaxies is the degeneracy of the solution. In
fact, if the fit is doing blindly, the algorithm would try, in some cases,
to accommodate the exponential component fitting the inner regions of the
profile. To avoid this our code selects an initial guess for the exponential
component that guaranteed that the outer
observable points are well described by this component in the first step of
the iteration. Even so, it could be possible that the final solution of the
fit would provide an answer where the inner points of the surface brightness
were described primarily by the exponential component. To avoid these cases,
the algorithm checks whether the central intensity of the fitted exponential
component is larger than the central intensity of the fitted S\'ersic
component. If this is the case the fit starts again changing the initial
conditions until a proper fit is achieved. This guarantees that the S\'ersic
component is always describing the inner regions of the profile and that 
$B/T$ is meaningful.

\begin{figure}
\begin{center}
    \epsfig{file=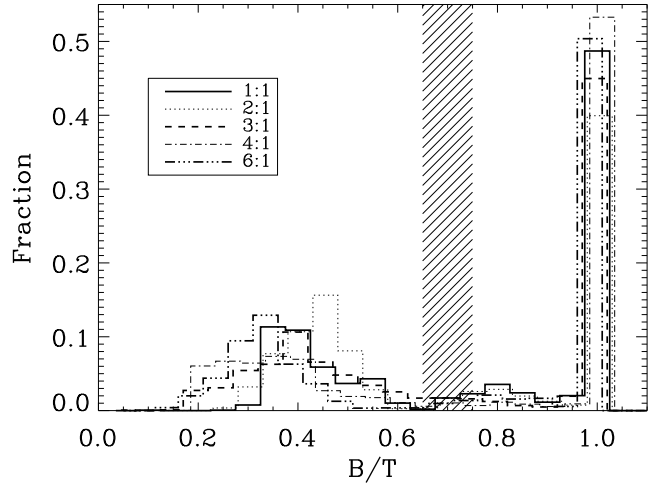,width=0.5\textwidth}
  \caption{Histogram of the bulge--to--total ratios ($B/T$) for all projected
  1:1, 2:1, 3:1, 4:1, and 6:1 merger remnants. The shaded area indicates
  the transformation region around $B/T \approx 0.7$ between disc+bulge
  systems and pure bulge systems. The curves are slightly shifted for better 
visibility.\label{bulge2total_hist_allinone}}     
\end{center}
\end{figure}

\begin{figure*}
\begin{center}
    \epsfig{file=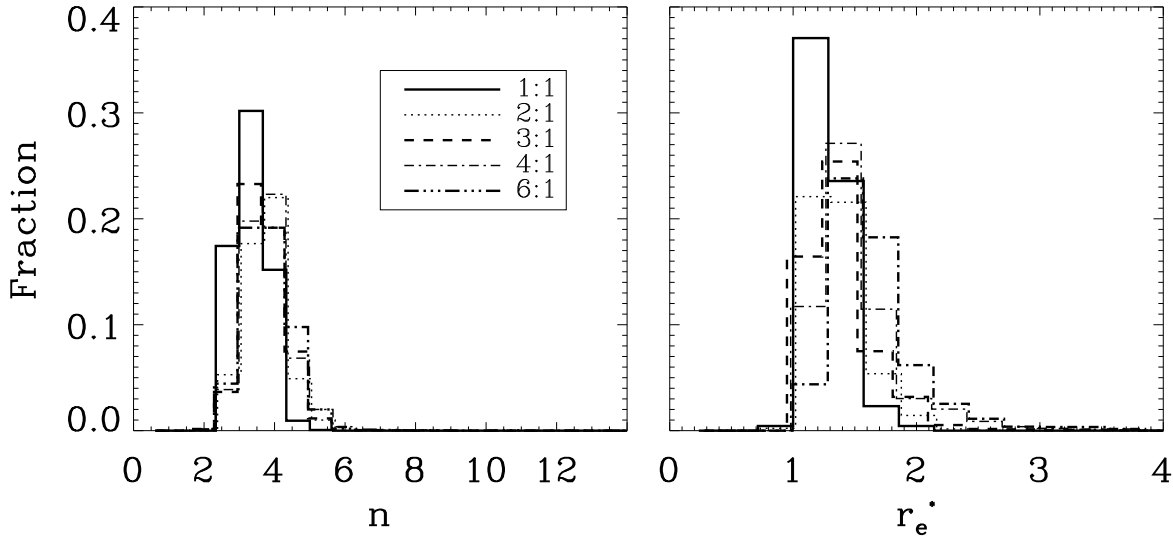,width=0.9\textwidth}
  \caption{Distribution of properties for all projections
  classified as pure bulges for the projected 1:1, 2:1, 3:1, 4:1, and 6:1  
merger remnants. The data are shown as a fraction of all projected remnants at a given 
mass ratio. {\it Left}: The S\'ersic-index n. The distributions show a weak dependence on
  mass-ratio and peak at around $3<n<4$.  {\it Right}: Effective radii which peak at 
$1 < r_{e}^* < 1.7$ for all mass  
  ratios. {\it Right}: \label{b_allinone}}  
\end{center}
\end{figure*}

\begin{figure}
\begin{center}
    \epsfig{file=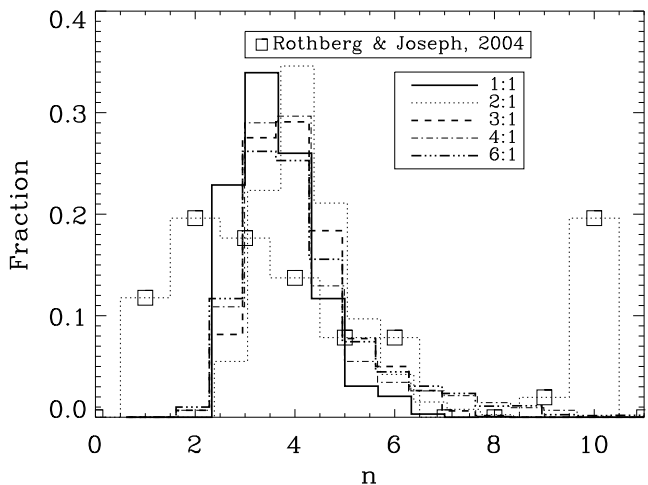,width=0.5\textwidth}
  \caption{Histogram of the S\'ersic-index n for all projections. 
In contrast to Fig.\ref{b_allinone} we have plotted all results from the 
S\'ersic only fit independent of the classification (e.g. $B/T$) of the projection. 
The distributions peak at similar values but show a tail towards larger
values of $n$. The observed (normalised) distribution of local merger remnants 
\citep{2004AJ....128.2098R} is  
indicated by open squares. 
\label{ns_hist_allinone_simple}}  
\end{center}
\end{figure}

\section{Statistical analysis}
\label{STATISTIC}

\begin{table}
 \centering
 \begin{minipage}{70mm}
  \caption{Photometric properties (see Figs.  \ref{bulge2total_hist_allinone} - \ref{db_allinone}) 
for mergers with bulges for different 
mass ratios with $1\sigma$ deviations.\label{table1} 
}
  \begin{tabular}{@{}ccccccc@{}}
  \hline
  Mass ratio  & 1:1 & 2:1& 3:1 & 4:1 & 6:1\\
 \hline
          n & 3.1 & 3.56 & 3.60 & 3.66 & 3.66 \\
 $\pm$  & 0.51 & 0.59 & 0.62 & 0.66 & 0.72 \\ 
\hline
      $r_e^*$  & 1.19 & 1.27 & 1.36 & 1.47 & 1.60 \\
$\pm$& 0.16 & 0.23 & 0.31 & 0.43 & 0.51 \\    
\hline
\hline
      $r_{eG}$  & 1.15 & 1.09 & 1.11 & 1.16 & 1.20 \\
$\pm$& 0.08 & 0.09 & 0.10 & 0.08 & 0.1 \\    
\hline
      $n_{B}$  & 1.70 & 1.70 & 1.66 & 1.39 & 1.32 \\
$\pm$& 0.33 & 0.25 & 0.56 & 0.41 & 0.48 \\    
\hline
      $r_{eB}$  & 0.42 & 0.38 & 0.39 & 0.33 & 0.33 \\
$\pm$         & 0.09 & 0.07 & 0.19 & 0.11 & 0.15 \\    
\hline
      $h_{D}$  & 1.32 & 1.29 & 1.22 & 1.18 & 1.19 \\
$\pm$         & 0.24 & 0.20 & 0.29 & 0.29 & 0.18 \\    
\hline
\end{tabular}
\end{minipage}
\end{table}
In the appendix we show a detailed analysis of the initial conditions and individual
examples of merger remnants. As we show there the properties of the merger
remnants depend strongly on the respective projection angle under which the
remnant would be observed. In this section we present the results of a
statistical analysis of all merger remnants, with and without bulges in the
progenitor discs, by analysing 100 random projections of every merger
remnant grouped according to their mass ratios.  The results for mergers with bulges are summarized in 
Tab. \ref{table1}. 

\subsection{Progenitors with bulges}
\begin{figure*}
\begin{center}
    \epsfig{file=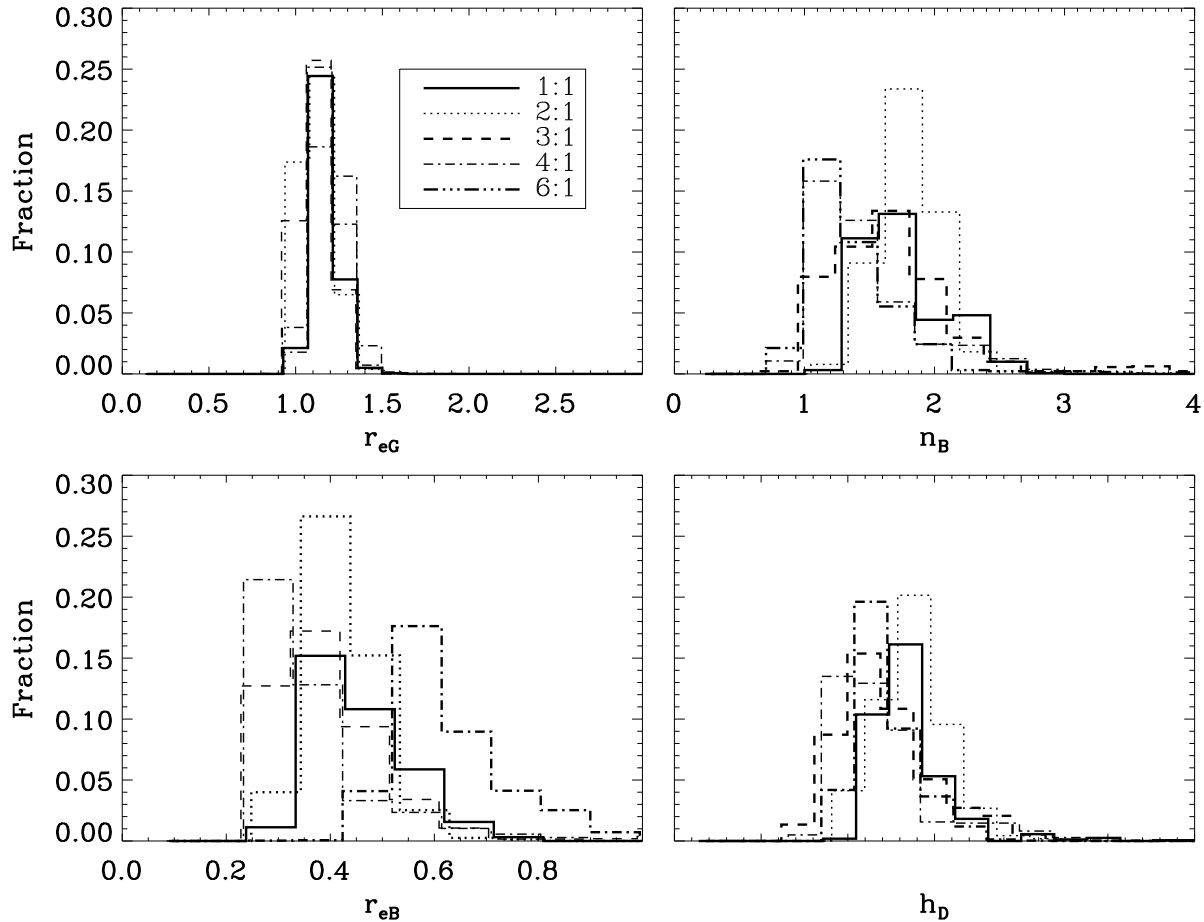,width=0.9\textwidth}
  \caption{Distribution of the properties for the projections classified as $D/B$ systems 
for all mass ratios. {\it Upper left}: effective sizes; {\it Upper right}: S\'ersic-indices 
of the bulge components; {\it Lower left}: Effective radii 
of the bulge components; {\it Lower right}: Scale lengths of the exponential components.}
\label{db_allinone}  
\end{center}
\end{figure*}

In Fig. \ref{bulge2total_hist_allinone} we show the bulge--to--total
($B/T$) ratios for all merger remnants with bulges and mass ratios of 1:1, 2:1,
3:1, 4:1, and 6:1. The transition region from bulge+disc systems to
pure bulge systems around $B/T =0.7$ is indicated by the shaded area. Small changes of
this threshold value would not influence the global results as all mass ratios show a 
minimum number of projections in the transition region. Fig. 
\ref{bulge2total_hist_allinone} shows that merging increases the
bulge fraction of the galaxy significantly. For comparison, the $B/T$
values of the initial disc do not exceed $B/T = 0.20$ (see appendix).  The $B/T$ ratios for every mass ratios cover 
a wide range from $0.2 < B/T <1$ and the distributions are remarkably similar. For every mass ratio
we would identify pure bulge systems as well as bulge+disc
systems. Using $B/T=0.7$ as the division line the percentage of pure
bulge systems would be
66\% for the 1:1, 51\% for the 2:1, 54\% for the 3:1, 58\% for the 4:1, and 58\% for 
the 6:1 remnants, respectively. 1:1 remnants have the largest number of 
projections which would be classified as pure bulges.
There is no trend for remnants with mass ratios of 2:1 and higher 
to be less dominated by bulge-only projections, e.g. 6:1
merger remnants do not have significantly more projections classified as disc+bulge
systems as e.g. 2:1 remnants. However, there is a trend that high mass-ratio remnants 
classified as $BD$ can have lower $B/T$ indicating that the exponential component 
is becoming more dominant.   

\begin{figure}
\begin{center}
    \epsfig{file=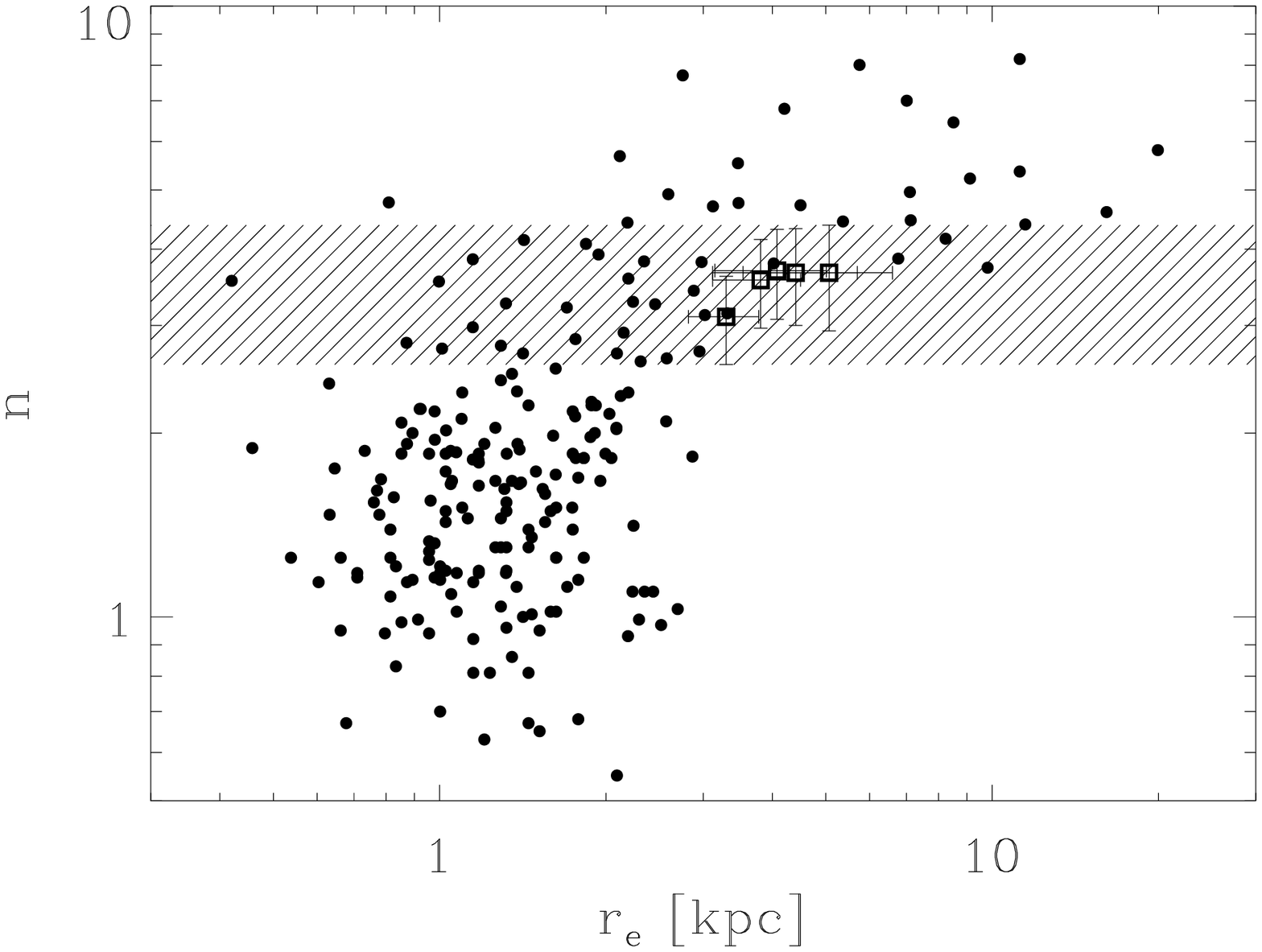,width=0.5\textwidth}
  \caption{Distribution of effective radius $r_e^*$ versus shape parameter $n$ for the simulated merger 
remnants at different mass ratios (symbols with error bars from Tab. \ref{table1})and a sample of 
observed (dots) elliptical galaxies \citep{1990AJ....100.1091P,1990A&AS...86..429C,1994A&AS..106..199C,1998A&A...333...17B,2004ApJ...602..664G}. We have assumed an initial disc scale length of 3kpc for a good agreement with the data. 
However, the simulations are scale free and can be shifted horizontally along the shaded area.   
\label{re_vs_n_obs}}  
\end{center}
\end{figure}

In the following we focus on the properties of the projections
classified as pure bulge systems with $B/T > 0.7$ (the distributions 
in Figs. \ref{bulge2total_hist_allinone} - \ref{MCN_allinone} 
are shown as fractions of all projected 
remnants at a given mass ratio). The distribution of
the best fitting S\'ersic-indices for the five mass ratios is shown in 
Fig. \ref{b_allinone}. The distributions are very 
similar for all mass ratios and peak between $3 < n < 4$. The
bulk of the projections lie in the range $2.5 < n < 6$ which is in
general in good agreement with observed values of intermediate mass 
giant elliptical galaxies (e.g. \citealp{2004ApJ...601L..33V,2004ApJ...600L..39T,
2004MNRAS.355.1155D}, and references therein). Ellipticals with $M_B > -18$ have on average 
smaller shape parameters, massive ellipticals with $M_B < -20$ and large 
velocity dispersions have on average shape parameters $n > 4$. This is demonstrated in
 Fig. \ref{re_vs_n_obs} where show the mean values of the shape parameters and 
the effective radii in comparison to a sample of elliptical galaxies 
\citep{1990AJ....100.1091P,1990A&AS...86..429C,1994A&AS..106..199C,1998A&A...333...17B,2004ApJ...602..664G}. 
The numerical 
simulations are scale free and can therefore be shifted to smaller as well as larger sizes 
along the shaded region. We have assumed a scaling of 3kpc for the initial disc scale length
 which is in best agreement with observations. Ellipticals with larger shape (smaller) 
parameters which are in general also more massive (less massive) can not be fitted by our 
merger remnants.  Interestingly, our distributions of the shape is very similar to the distributions 
of dark matter halos in cosmological simulations \citep{2005ApJ...624L..85M}.  

For a better comparison with the sample of local merger remnants 
by \citet{2004AJ....128.2098R} who fitted only S\'ersic profiles 
we show in Fig. \ref{ns_hist_allinone_simple} the distribution of the indices 
for the S\'ersic only fit for every projection independent of its $B/T$ value. 
The simulated distributions peak at similar values and
 now have a prominent tail towards larger values of $n$. The distribution 
is similar to the local merger remnants from \citet{2004AJ....128.2098R}. However, we 
are missing all projections with $n \le 1$ and the observed distribution peaks at lower 
$n$ ($n=2$). This might reflect the fact that some progenitors had less massive bulges
resulting in smaller values of $n$ (see Section \ref{BULGELESS}). We also
find indices larger than $n=8$ for mass ratios 4:1 and 6:1. 
\citet{2004AJ....128.2098R} have suggested that these 
values result from excess light at the center. In our case it is those remnants 
that show a concentrated bulge plus a very clear exponential component and were classified
as two-component systems. 

Comparing to previous simulations \citet{2005MNRAS.357..753G} have found 
values between $3 < n < 8$ (they only performed S\'ersic-fits) for a small set 
of 1:1, 2:1 and 3:1 remnants which is in good agreement with our results 
in this mass-ratio range taking into account that their initial bulge was more massive 
(the bulge-to-disc ratio was 1/2). In addition, 
they also found no dependence of $n$ on the mass-ratio.

The distribution of the effective radii for the systems classified as
pure bulges $B/T > 0.7$ is shown in Fig. \ref{b_allinone}. To correct for the 
flattening we computed the circularised effective radius 
$r_e^* =r_e \sqrt{1 - \epsilon}$ using the semi-major axis effective radius $r_e$ and 
the ellipticity, $\epsilon$, of the image. The distributions for all mass-ratios 
are similar and peak at $1 < r_{e}^* < 1.7$ 
with a tail extending to values as large as $r_e^*= 3$. 
Compared to the initial conditions the photometric sizes do, on average, not 
or only weakly increase. This result seems at first surprising, however it was already 
found by \citet{1993ApJ...409..548H} that the half-mass radius of the remnant can 
be of a similar size than the initial disc. For a direct comparison we have computed the 
projected half mass radius of one of our runs that is identical to "1-nrb" in 
\citet{1993ApJ...409..548H}. In three different projections our values (1.80,1.2,1.15) 
are similar to theirs (1.92, 1.31, 1.18) taking into account that we have 4-5 times more 
particles for the luminous components and 7 times more particles in the halo component.  

We will now focus on all projected remnants that are classified as
disc+bulge systems (Fig. \ref{db_allinone}). In the upper left we
show the distribution of the global effective radii as they have been calculated 
using Eqn. \ref{re_global}. All remnants classified as $DB$ systems have similar 
sizes of $r_{eG} \approx 1.2$, similar to the initial conditions. Consequently, 
independent of $B/T$ and the initial mass-ratio all remnants have similar sizes
than the initial discs. The distribution of the S\'ersic-indices and the
effective radii of the bulge components of the BD fits are shown in
the upper right of Fig. \ref{db_allinone}. The
S\'ersic-indices for the bulge components of 1:1, 2:1 and 3:1 remnants peak
at values  of $n_B = 1.8$. For 4:1 and 6:1 remnants the distributions
peak at slightly lower values around $n_B = 1.1$ with tail of 
higher values extending to $n_B = 2.8$. 
The effective radii peak around $0.2 < r_{eB} < 0.8$ for all mass-ratios. 
The maximum  values are about a factor four larger than the effective radius 
of the bulge component when analysed for the initial conditions. 
The scale lengths of the fitted disc components peak around $1.2 < h_D < 1.3$ 
which is about 30\% larger than the initial disc scale lengths (Fig. \ref{db_allinone})

\subsection{Bulge-less progenitors}
\begin{figure}
\begin{center}
    \epsfig{file=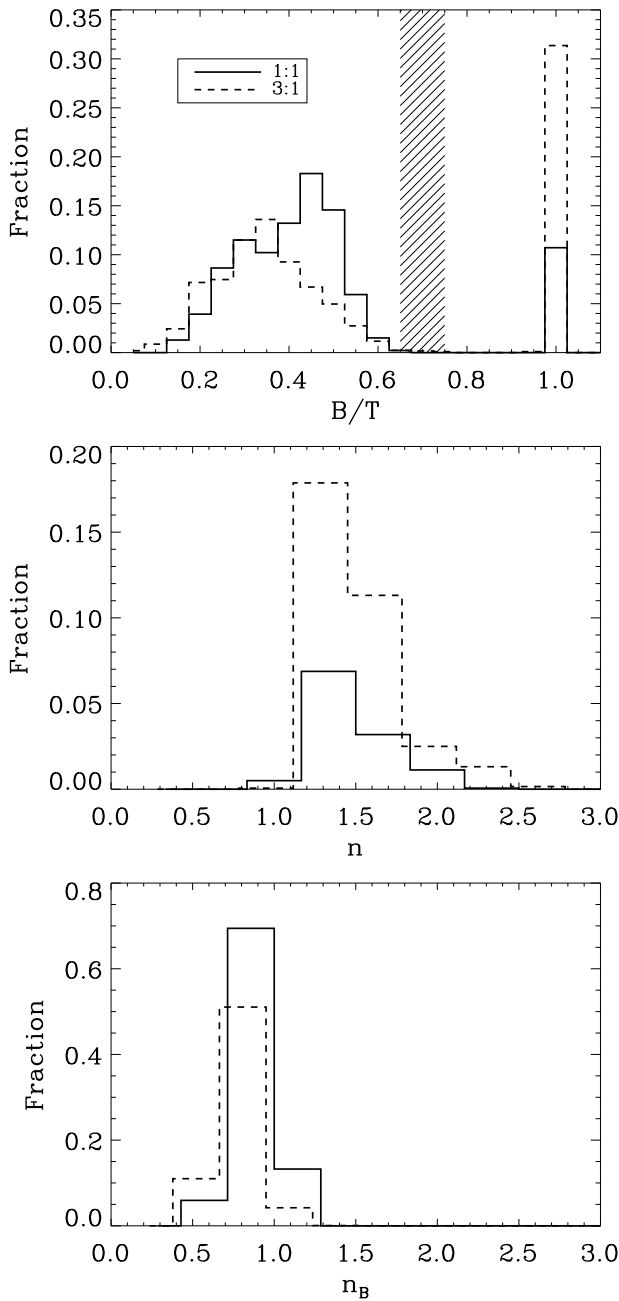,width=0.5\textwidth} 
  \caption{{\it Upper panel}: Histogram of the bulge--to--total ratio (B/T) for all projected
  1:1 and 3:1 bulge-less  merger remnants. The shaded area around  $B/T = 0.7$ divides 
bulge + disc systems from pure bulge systems. {\it Middle panel}: Distribution of the S\'ersic index for all projections
  classified as pure bulges for the projected 1:1 and 3:1 
  merger remnants without bulges. {\it Lower panel}: Distribution of the S\'ersic index of the bulge components  
    for the disc+bulge fits for the mass
    ratios 1:1 and 3:1. \label{MCN_allinone}}    
\end{center}
\end{figure}

\begin{figure}
\begin{center}
    \epsfig{file=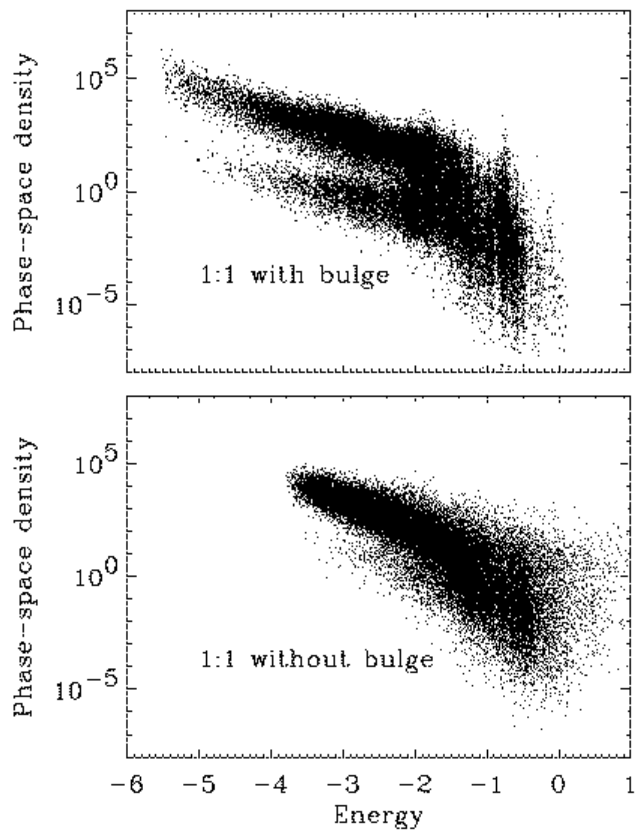,width=0.5\textwidth}
  \caption{Phase-space densities for typical 1:1 merger remnant with and without 
a bulge in the progenitor disc. The merger without the bulge has a lower phase-space density.
 \label{phase_11MCS_8}}      
\end{center}
\end{figure}

\label{BULGELESS}
We have re-simulated the mergers with mass ratios 1:1 and 3:1 using
progenitor discs without bulge components to investigate the influence
of the bulge on the surface density profile of the merger remnant.

In the upper panel of Fig. \ref{MCN_allinone} we show the result for the 
distribution of $B/T$-ratios of the projected remnants. The result is significantly
different to the analysis of the remnants with bulge (see
Fig. \ref{bulge2total_hist_allinone}). Most projected remnants
are classified as two-component systems. The distributions peak 
at $B/T = 0.5$ for the equal-mass remnants and $B/T = 0.35$ for the 3:1 remnants, 
almost reflecting the mass-ratios of the progenitor
discs. From the lower two panels in Fig. \ref{MCN_allinone} 
it becomes clear that the systems classified as pure bulges have $1.2 < n < 2.5$ 
and therefore very shallow surface density profiles and, if the system is classified 
as a two component system, the S\'ersic index of the bulge is smaller than one. 

Interestingly, S\'ersic indices as small as $n=1$ have been observed for local merger remnants 
\citep{2004AJ....128.2098R} which might indicate that some progenitor discs did not have 
prominent stellar bulges (or bulges with low concentration). 
 
For a similar set of simulations \citet{2005MNRAS.357..753G} fitting a S\'ersic only 
model for all remnants find a range of $2.4 < n < 3.2$. 
There might be two origins for this discrepancy: all our remnants that appear more concentrated 
show a two component structure and were not classified as one component systems. If we 
limit ourselves to S\'ersic fits only like  \citet{2005MNRAS.357..753G} the indices would 
be in the range $1.2 < n < 3.1$. The fact that we find smaller values for $n$ might be caused 
by the different choice of initial conditions for the discs. \citet{2005MNRAS.357..753G} use 
a truncated exponential disc which leads to remnant profiles that can not be fitted properly 
in the outer parts. Our remnants show very exponential profiles in the outer parts which 
are fitted with small residuals leading to smaller over all values of $n$. 
Despite the differences in detail the results of \citet{2005MNRAS.357..753G} and 
our results clearly indicate that collisionless merger remnants of pure 
disc systems do not evolve into concentrated systems with surface density profiles 
similar to giant elliptical galaxies. Mergers without bulges also have lower phase-space 
densities as we show for a typical example in Fig. \ref{phase_11MCS_8}. The remnant with bulge 
has more strongly bound particles and the maximum phase space density is at least a factor 10 
larger than for the bulge-less remnant. The phase-space densities have been computed as in 
\citet{2005MNRAS.356..872A} using the program kindly provided by the authors. This has been suggested  
for a long time \citep{1980ComAp...8..177O,1986ApJ...310..593C,
1993ApJ...409..548H} and has to be taken into account when interpreting simulations
of stellar mergers of pure disc galaxies. In the following we only consider properties of 
remnants of mergers with bulges in the progenitor discs.

\section{Global parameter relations} 
\label{GLOBAL}
\begin{figure}
\begin{center}
    \epsfig{file=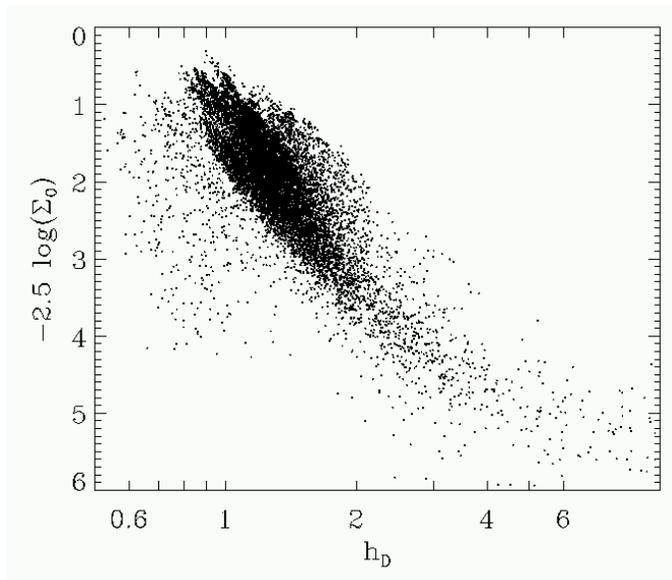,width=0.5\textwidth}
  \caption{Disc scale length, $h_D$, versus the central surface density of the disc, 
$\Sigma_0$, for all remnants with bulges and all mass ratios.  \label{h_vs_I0_allinone}}    
\end{center}
\end{figure}

\begin{figure}
\begin{center}
    \epsfig{file=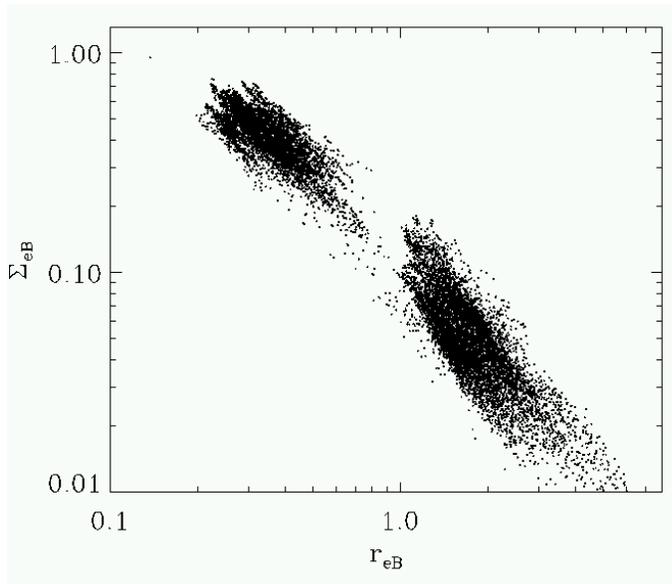,width=0.5\textwidth}
  \caption{Effective radius of the bulge $r_{eB}$ versus the mean effective surface density of 
the bulge, $\Sigma_{eB}$, for all projected remnants with bulges. If the galaxy was classified as a 
one component system we assumed $r_{eB} = r_e$. 
\label{re_vs_Ie_allinone}}    
\end{center}
\end{figure}

\begin{figure}
\begin{center}
    \epsfig{file=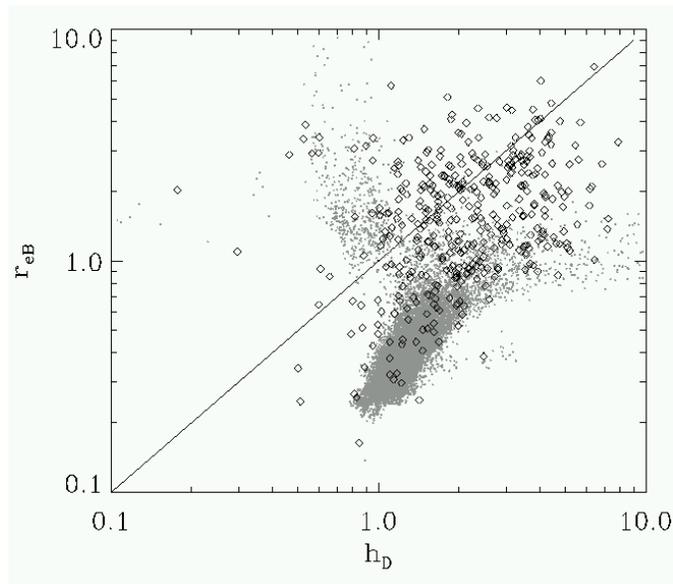,width=0.5\textwidth}
  \caption{Disc scale length, $h_D$, versus the effective radius of the bulge, $r_eB$, for all 
projections (grey dots) assuming the best fitting disk+bulge model. Observational data for E/S0 galaxies 
from 
\citet{2004MNRAS.355.1155D} are overplotted (open squares) and scaled down by a factor of 3 
(same scaling as in Fig. \ref{re_vs_n_obs}) corresponding to an initial model disk scale length of 3kpc.    \label{h_vs_re_allinone}}    
\end{center}
\end{figure}

In Figs. \ref{h_vs_I0_allinone} and \ref{re_vs_Ie_allinone}
we show the correlation between the disc scale length and the central surface density of the disc 
as well as the effective radius and the mean effective surface density of the bulge. In both cases 
there is a reasonably tight correlation in the sense that larger systems have lower surface 
densities. The bimodal distribution in Fig. \ref{re_vs_Ie_allinone} arises from the fact that 
there is no continuous transition between the systems classified as DB systems (upper left) 
and systems classified as one-component bulges (lower right). The trend for the correlations 
is surprisingly similar to observed galaxies (see e.g. \citealp{2004MNRAS.355.1155D}; their Figs. 6
and 7). The correlation between the bulge effective radii and the disc scale lengths is 
shown by the grey dots in Fig. \ref{h_vs_re_allinone}. 
Systems with larger bulges also have larger discs. On average $h_D$ is a factor of 3-5 
larger than $r_{eB}$. In contrast, \citet{2004MNRAS.355.1155D} (open squares in Fig. 
\ref{h_vs_re_allinone}; here we have scaled the obervational data down by a factor 3 assuming 
 an initial disk scale length of 3kpc for the model disc) find that bulges 
in observed early--type galaxies are on average only a factor of 2 smaller than the discs. 
This result might reflect the fixed $B/T$ ratio in our initial conditions of our sample as we 
find a good agreement with observations for small bulge sizes. 
The few projections 
that are found to have similar sizes in both components also have the largest $B/T$ values 
$(0.7 < B/T < 0.9)$. As we have seen in Fig. \ref{bulge2total_hist_allinone} there are 
only a few projections in this regime. 

\begin{figure}
\begin{center}
    \epsfig{file=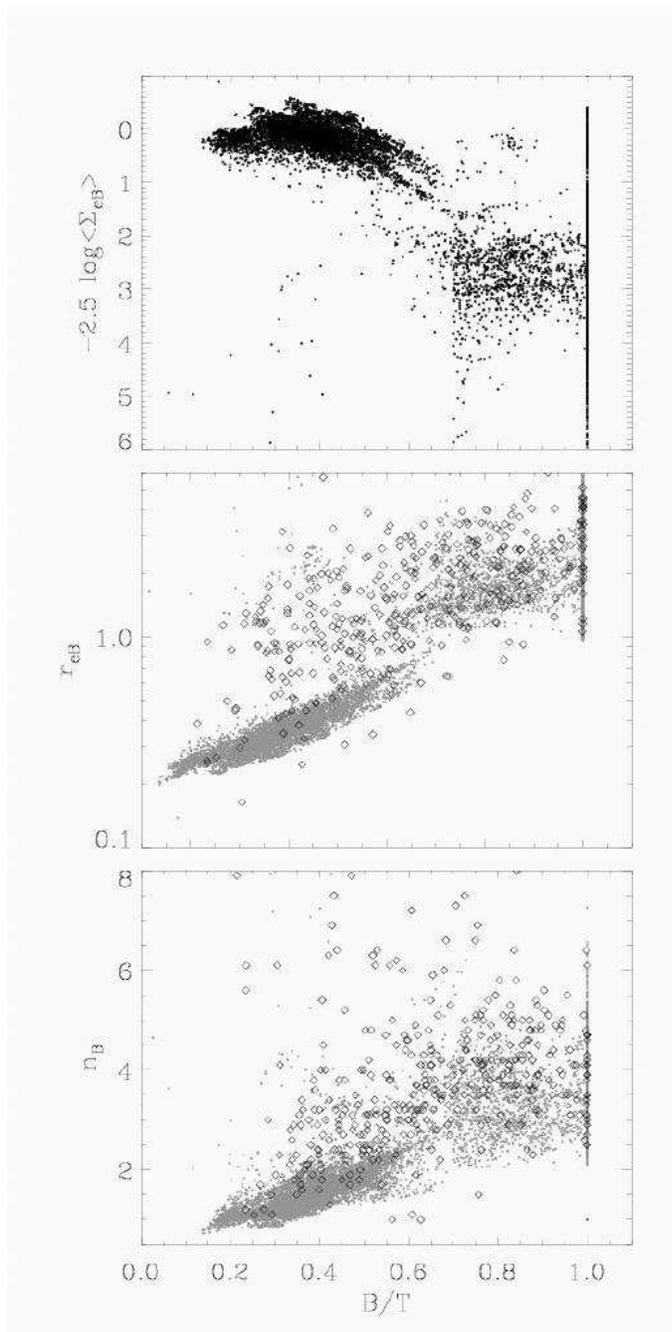,width=0.5\textwidth}
  \caption{{\it Upper panel}: Mean effective surface density of the bulge versus 
$B/T$. {\it Middle panel}: Effective radius of the bulge versus $B/T$. 
For systems with $B/T=1$ we use $r_{eB} = r_{e}$. {\it Lower panel}:
S\'ersic index of the bulge versus $B/T$. For systems with $B/T=1$ we use $n_B = n$. 
We show all projected remnants with bulges in the initial condition discs. Observational data 
for E/S0 galaxies from \citet{2004MNRAS.355.1155D} are overplotted (open squares) for the two lower panels and 
scaled down for $r_{eB}$ by a factor of 3 (same scaling as in Fig. \ref{re_vs_n_obs}) which corresponds to an initial disk scale length of 3kpc. \label{bulge2total_correls}}    
\end{center}
\end{figure}

In Fig. \ref{bulge2total_correls} we show, for every projection and all mass ratios,  $B/T$ versus
the mean effective surface density, $<\Sigma_{eB}>$, the effective radius $r_eB$ and the 
S\'ersic index of the bulge component $n_B$. The points on the vertical line, 
$B/T = 1$, indicate the projected remnants classified as pure bulges. Observations from 
\citet{2004MNRAS.355.1155D} are overplotted as open diamonds. 
There is a weak trend for galaxies with a larger $B/T$ to have
a smaller bulge effective  surface density (upper panel of Fig. \ref{bulge2total_correls}). In
addition, galaxies  with smaller $B/T$ ratios also have bulges with smaller effective radii 
(middle panel of Fig. \ref{bulge2total_correls}). There is a good general agreement with 
observations but still the bulk of the simulated projections have small bulge sizes. 
There is also a correlation between  $B/T$ and
the S\'ersic index of the bulge. Remnants  with $n_B \approx 1$ have small values of $B/T \approx
0.3$ and $n_B$ is increasing with  increasieng $B/T$. A similar correlation has beeen found by 
\citet{2004MNRAS.355.1155D} (open diamonds) for their  large sample of observed early--type
galaxies.  However, the observed  values for $n_B$ have a larger spread and appear to be on average
20-30\% larger for $B/T < 1$. The correlations for $r_{eB}$ and $n_B$ are much tighter for $B/T <
0.7$. In this regime  they are also much tighter than the observed correlations
\citep{2004MNRAS.355.1155D} which might  be due to the fixed bulge-to-total ratio in the initial
conditions. We found no other significant  correlation of the profile properties with global
properties of the remnants like anisotropy,  isophotal shape, or shape of the line-of-sight
velocity distribution. In particular, there  is no correlation between the projected central
velocity dispersion of the remnants (which varies  from a mean value of $\sigma = 0.5$ for 6:1
remnants to a mean value of $\sigma = 0.68$ for 1:1 remnants  in computational units) and the
S\'ersic index of the bulge. However, this correlation has been  found to be rather tight for
observed ellipticals \citep{2001AJ....122.1707G,2004ApJ...601L..33V}. As our simulations are scale
free we can make the simple experiment and try to shift the velocity scale of the remnants to the
region  where the observed $n$ shows a similar distribution peaked around $n=3.5$ . Using the
\citet{2004ApJ...601L..33V}  data this would correspond to a small velocity range of $130km/s \le
\sigma \le 170km/s$.  Galaxies with lower $\sigma$  have smaller $n$, those with higher dispersion
tend to have larger $n$.  This would shift our more massive initial disc galaxy to the Milky Way
regime. Taking the above considerations at face value we might conclude that only binary mergers of
evolved early--type spiral galaxies with masses similar to the Milky Way contribute to the
intermediate mass population of elliptical galaxies. Turning the  argument around we can almost
certainly exclude that the most massive ellipticals are made by binary  mergers of disc galaxies.
This is in agreement with the kinematical analysis of \citet{2003ApJ...597..893N}.

\section{Is the exponential outer component a real disc?} 
\label{NATURE}
\begin{figure}
\begin{center}
    \epsfig{file=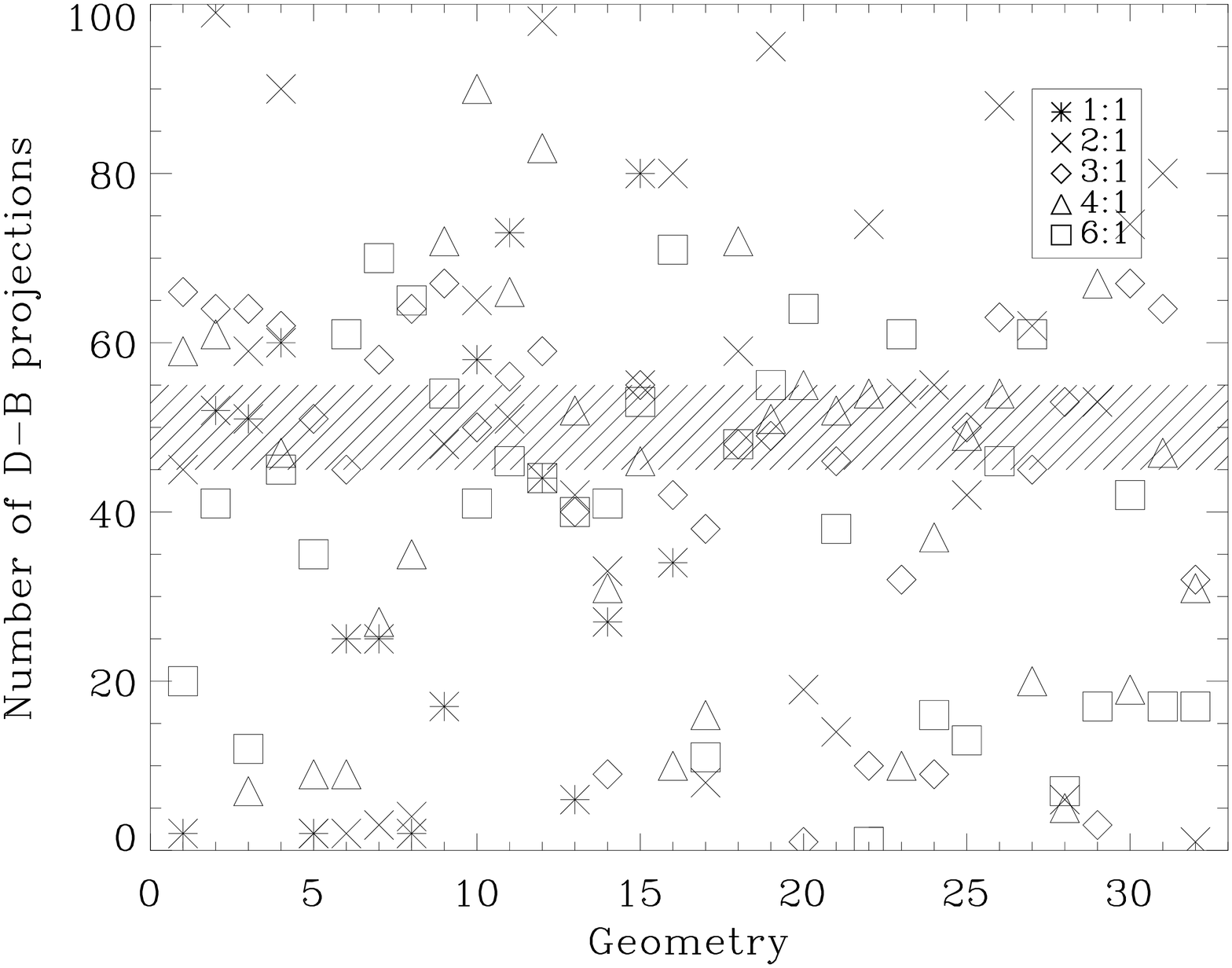,width=0.5\textwidth}
  \caption{Number of projections (out of 100) classified as DB systems for mass ratios 1:1-6:1.
The numbers on the abscissa indicate individual merger geometries with initial disc orientations as in 
\citet{2003ApJ...597..893N}, Table 1. 
\label{ns_vs_geo_allinone}}    
\end{center}
\end{figure}
\begin{figure}
\begin{center}
    \epsfig{file=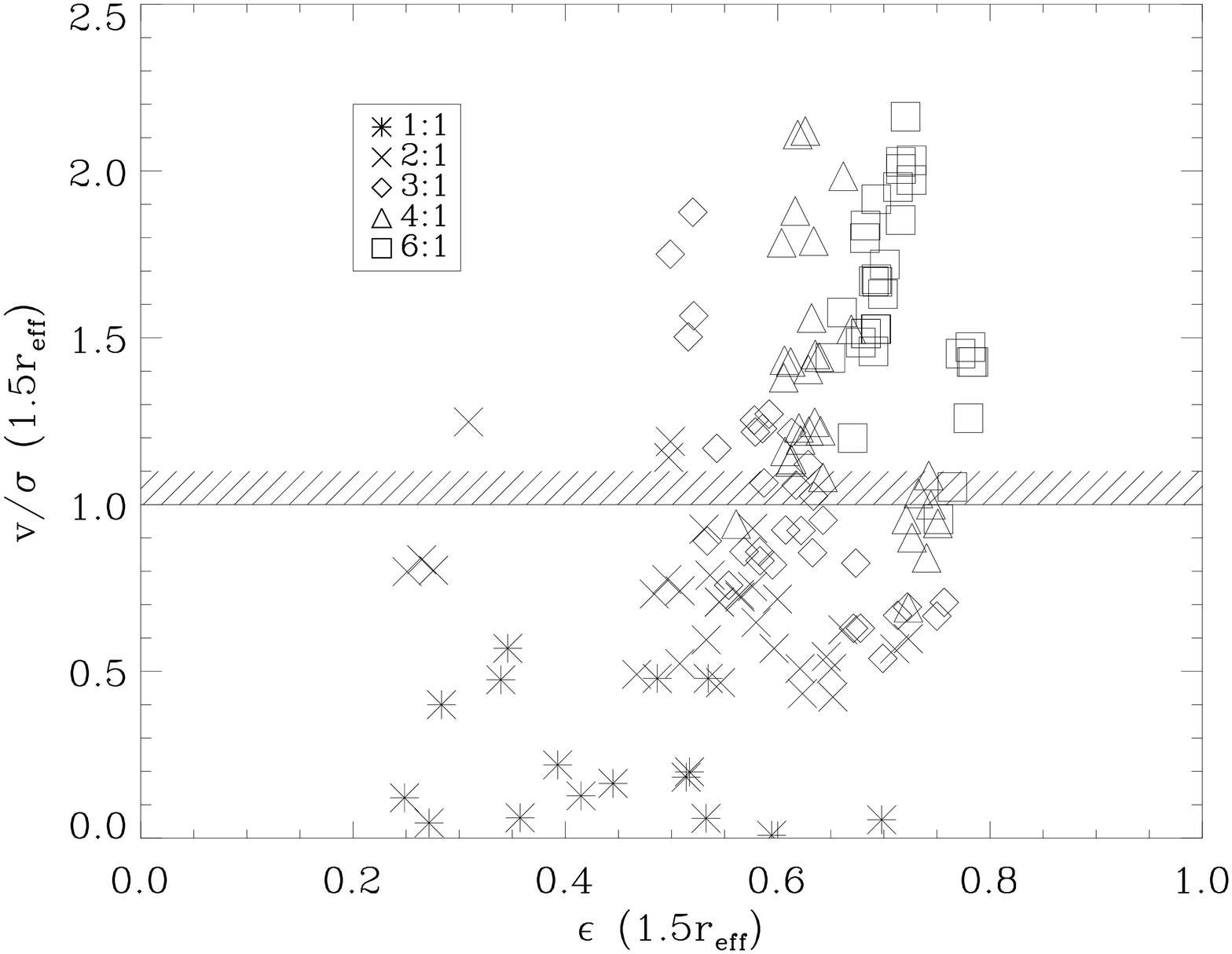,width=0.5\textwidth}
  \caption{Local ratio of $v/\sigma$ versus $\epsilon$ at $1.5 r_e$ for the edge-on projections
of all merger remnants. We assume disc-like properties if $\epsilon \gtrsim 0.65$ and 
$v/\sigma \gtrsim 1$.
\label{ellmax_vs_vsmax_allinone}}    
\end{center}
\end{figure}

In Fig. \ref{ns_vs_geo_allinone} we show the fraction of the 100 random 
projections of every remnant that were classified as a two-component systems
(the rest was classified as one component systems) as a function of the initial geometry of the
mergers. Only for a few cases, mostly for mass 
ratios 1:1 and 2:1, the remnant properties are independent of projection effects. 
Statistically we find a weak influence of the initial mass ratio or the initial geometry 
alone on the classification of the remnants. Typically the changes in classification are caused 
by a combination of different mass-ratios, initial geometries and projection effects. However, 
the presence of an outer exponential is a general feature which 
is common to most remnants. But what is the nature of this outer component? Naively one would 
assume that it represents a classical  exponential disc. At least for 1:1 and 2:1 remnants this interpretation is 
questionable as e.g. \citet{2003ApJ...597..893N} have demonstrated that the kinematical 
properties of these remnants are very similar to slow-rotating hot stellar systems. Interestingly, 
if investigated in detail remnants of all mass ratios show similar degrees 
of anisotropy and the main kinematical difference appears to be the amount of rotational support 
\citep{2005astro.ph..4387B,2005astro.ph..4595B}. On the other hand, the material in the outer regions 
of the remnants originates from the exponential disc component of the progenitor discs and the exponential profile
of the remnants could point to this origin. 

If the exponential component of the remnants was an outer disc it should be significantly flattened with a high 
ellipticity ($\epsilon \gtrsim 0.7$) if seen edge-on. Dynamically it should be significantly supported by 
rotation with the local ratio of the line-of-sight velocity and the velocity dispersion, $v/\sigma$, greater than 
unity. Proper discs in ellipticals might have significantly larger values than that 
\citep{1999ApJ...513L..25R}, however there also exist S0 galaxies which have confirmed 
stellar disc components but $v/\sigma < 1$ 
\citep{1992ApJ...400L...5R,1992ApJ...394L...9R}. In case of NGC 4550 this is caused by two 
counter-rotating discs which could only have been detected by detail kinematical analysis. In this sense 
the ellipticity might be the stronger constraint for our first order analysis. 

To test the properties of the exponential components of the remnants we measure $\epsilon$ and  $v/\sigma$
at a radius of $1.5 r_{eG}$ were the exponential component typically becomes dominant. Typical bulges in our 
remnants have sizes of $r_{eB} < 0.5 r_{eG}$ (Fig. \ref{db_allinone}) and dominate 
only the center.  In Fig. \ref{ellmax_vs_vsmax_allinone} we show the 
local values of $\epsilon_{1.5}$ and $v/\sigma_{1.5}$ at $1.5 r_e$ for every remnant seen along the 
intermediate axis (edge-on). 1:1 and 2:1 remnants do not show high ellipticities and 
are dynamically hot. In this case the exponential outer component can not be associated with a
disc. Its properties are more similar to hot outer halos of cD galaxies \citep{1996ApJ...465..534G}. 
Remnants with mass ratios of 6:1 on the other hand show very high ellipticities, $\epsilon_{1.5} \gtrsim 0.6$, 
and in general $v/\sigma_{1.5} > 1$, in some cases even $v/\sigma_{1.5} > 2$ which is a strong 
indication for the presence of an intact disc component that was heated during the merger.
For comparison: the initial condition disc has $\epsilon_{1.5} = 0.85$ and $v/\sigma_{1.5} \approx 5$.  
3:1 and 4:1 remnants have intermediate properties for which the two-component projections 
are in good agreement with the observed ellipticals with exponential outer profiles and 
elliptical-like kinematics as characterized by $v/\sigma \simeq 1$ \citep{2002A&A...393L..89J}. 
As proposed by \citet{1998ApJ...502L.133B} 6:1 and 4:1 remnants are good candidates for S0 galaxies 
(see also \citealp{2004A&A...418L..27B}). They show strong evidence for a heated disc 
component but still elliptical like surface density properties in some projections indicating that depending on the 
projection the remnants would be classified as S0 or elliptical galaxies. We do not find outer components with 
a flattening smaller than $\epsilon =0.2$ which is probably due to the limited heating during the interaction 
process. Note that only the strongly interacting mergers with mass ratios of 1:1 and 2:1 lead to values of 
$\epsilon  < 0.45$

\citet{2004A&A...418L..27B,2005A&A...437...69B} have analysed disc merger simulations
including star formation and argue that ellipticals with outer exponentials can be produced by mergers 
with mass-ratios in the range 4.5:1 to 7:1. Based 
on our larger statistical sample (however, without star formation) we 
conclude that all mass ratios can lead to remnants with outer exponential profiles and 
elliptical-like kinematics.  
We can think of several possible reasons for the subtle differences in the conclusions: either 
the presence of a dissipative component and star formation changes the properties of 
mergers with mass ratios of 1:1 to 3:1 or \citet{2005A&A...437...69B} have missed the 
exponential components due to their simple classification criterion. In addition, 
\citet{2005A&A...437...69B} use different initial conditions and a different numerical 
algorithm (grid code) with a spatial resolution that is a factor of two lower than for our simulations. 
Taking all these differences into account the general behaviour of remnants with mass ratios larger 
than 3:1 is in good agreement. 

The orbital structure of the remnants provides additional evidence that the structure 
of the exponential outer components change with mass ratio. \citet{2005MNRAS.360.1185J} have shown 
that at radii larger than the effective radius 1:1 remnants have equal amounts 
of box (box+boxlet) and tube orbits. In some cases major-axis tubes are also populated to a similar 
degree as minor-axis tubes supporting a round shape. On the other hand, 3:1 and 4:1 remnants are, at larger 
radii, clearly dominated by tube orbits. Tube orbits carry the angular momentum in the systems and are 
the typical orbits in stellar discs.

\section{Summary \& Discussion}
\label{CONCLUSIONS}

The surface density profiles of an unbiased sample of simulated 
stellar merger remnants of disc galaxies with bulges (mass ratios 1:1, 2:1, 3:1, 
4:1, and 6:1) and without bulges (mass ratios
1:1 and 3:1) in the progenitor discs have been investigated in detail. Every remnant has
been analysed from 100 random viewing angles resulting in 19200 projected remnants in total. 
We have fitted the profiles using a single S\'ersic function, and a S\'ersic function 
plus an exponential.  Based on the resulting bulge--to--total
ratios --- systems with high bulge--to--total ratios ($B/T > 0.7$) were considered as
one component --- and the quality of the fits we have classified the projected remnants either as 
one- or two-component systems.  In general the classification of every remnant changes 
with its projection and we find no strong correlation between the surface density properties of the 
remnants and the mass ratio of the initial discs.

Collisionless mergers of pure disc systems do not result in compact remnants similar to elliptical 
galaxies with large central phase space densities.  Most simulated remnants show a
two-component  structure with S\'ersic indices of the bulge $n_B < 1$.  In those cases where the
profile is well fitted with a single S\'ersic function the profiles are inconsistent  with $n > 2$ and
therefore inconsistent with properties of observed giant elliptical galaxies confirming earlier 
results \citep{1980ComAp...8..177O,1986ApJ...310..593C,1992ApJ...400..460H}. An additional gas 
component in the progenitor discs might be able to solve this problem. It does, however,  lead to a
concentrated peak of stars at the center unless very efficient feedback processes  are included
\citep{1996ApJ...464..641M,2005ApJ...620L..79S,2005astro.ph..3201C}. Interestingly, some observed
nearby merger remnants do show exponential K-band profiles  \citep{2004AJ....128.2098R} which
would indicate that their progenitors did not have  concentrated bulges. Whether those observed remnants
will evolve into bona-fide ellipticals  will depend on the amount of available gas and when and
where it is transformed into stars. Recently, \citet{2005astro.ph.11053R} showed that gas in disc mergers 
can change the central velocity dispersionsof the remnants and can contribute significantly to the tilt of
a the Fundamental Plane made by merger remnants. A full analysis of the surface density properties is, 
however, still missing.   

Merger remnants of discs with bulges cover a wide range of bulge--to--total 
ratios ($0.2 < B/T < 1$). At every mass-ratio more than 50\% of the projected remnants 
can be considered as one-component systems with a very similar S\'ersic index distributions 
in the range of $2< n < 6$. For all mass ratios the distributions peak in the range $3 < n < 4$. 
Statistically the photometric sizes of the remnants are of the same order than 
the size of the initial disc, independent of the mass ratio. 
In some projections the remnants can appear 50-100\% larger, in other projections the remnants 
can even appear smaller than the initial disc. Surprisingly, the difference 
in kinematic properties for different mass-ratios \citep{2003ApJ...597..893N} is more 
significant than the difference in the properties of the surface density profiles. We do not find 
correlations between the surface density profiles of our remnants and global properties 
like velocity dispersion or isophotal shape. As our models 
are scale free we can not explain the observed correlations between luminosity or velocity dispersion 
and shape index of the whole population of elliptical galaxies with our models alone. 
However, if our initial discs were scaled to a Milky Way-type galaxy the properties of the remnants 
would be consistent with intermediate mass early--type galaxies in the range of $-20 < M_B < -18$. 
This further supports the scenario that only part of the present day elliptical galaxy population 
can have formed by binary mergers of discs \citep{2003ApJ...597..893N}. More massive ellipticals 
are likely to have formed by a different process, e.g.early \citep{2005astro.ph.12235N} or late multiple mergers \citep{1996ApJ...460..101W} 
and/or mergers of early type galaxies ( e.g. \citealp{2005astro.ph..6425B,
2005MNRAS.361.1043G,2006ApJ...636L..81N}, and references therein). Of course the above 
conclusion is limited to our choice 
of initial conditions. Allowing for varying bulge-to-disc ratios and sizes for the initial discs with 
consistent bulge profiles might lead to remnants that follow the observed correlations. 
A first step in this direction has been attempted by \citet{2005MNRAS.360L..50A}. They did, however, 
not include bulge components in their initial discs which is problematic (see above) and a consistent 
study is therefore still missing. Furthermore we have been investigating the limiting case that 
the progenitor galaxies are gas free. It is not clear in how far the presence of gas and the uncertain 
physics of star formation and feedback will influence the surface density properties statistically. We 
can expect stronger changes for discs with higher gas fractions. Another study that has to be done 
in the future. 

More than 40\% of all remnants at all mass-ratios show clear signatures of an outer
exponential component. We have used additional local kinematic and photometric information at 
$1.5r_{eff}$ to determine the nature of this exponential component which is not possible 
by photometry alone. We found that the properties of the exponential component changes with mass ratio: 
For 1:1 and 2:1 remnants it resembles a dynamically hot spheroidal halo and for 6:1 remnants 
it resembles a flattened disc supported by rotation. 3:1 and 4:1 remnants have intermediate 
properties. It is known that the violent process during 1:1 and 2:1 mergers destroys 
the initial disc structure while higher mass ratio mergers are less violent and tend
to preserve the discs of the more massive progenitor 
\citep{2003ApJ...597..893N,2005MNRAS.357..753G,2005A&A...437...69B}. 
This interpretation is supported by the fact that 3:1 and 4:1 remnants are significantly 
more dominated by tube orbits in their outer parts than 1:1 remnants 
\citep{2005MNRAS.360.1185J}. We have shown that even collisionless 1:1 remnants, despite the 
violent merging, remember their initial state to some extend and can keep their outer 
exponential particle distribution. Observationally, \citet{2004MNRAS.355.1155D} find 
that most early-type galaxies show an exponential outer component if a two component fit is 
performed. The physical meaning of this finding is, however, still uncertain. We conclude that 
exponential outer components are a generic feature of binary disc mergers. They might directly 
indicate that mergers involving exponential stellar discs have played an important role during 
the formation of early-type galaxies. Of course there are other possibilities for the origin 
of outer exponential components. They might have formed from left-over gas after a disc merger 
\citep{2001ASPC..230..451N,2002MNRAS.333..481B,2005ApJ...622L...9S}, accretion 
of small satellites, accretion of gas from the halo followed by star formation etc. 
As all our 1:1-4:1 remnants have kinematic properties 
that are comparable to elliptical galaxies \citep{2003ApJ...597..893N} we can  
naturally explain observations of galaxies with elliptical-like kinematics but 
exponential-like profiles \citep{2002A&A...393L..89J}. In agreement with 
\citet{2004A&A...418L..27B} we find that the effect is strongest for high mass-ratios. 
In contrast to  \citet{2004A&A...418L..27B} we find that those systems can also 
form from mass ratios 1:1 - 4:1 and we do not see a sharp transition when changing 
the mass ratio from 4:1 to 6:1. 

We have found distinct correlations between the parameters from the surface density 
fits: (1) the sizes of the exponential component and the bulge are correlated; (2) a system with 
a larger bulge-to-total ratio has on average a larger S\'ersic index, a larger bulge effective 
radius and a lower effective surface density; (3) larger bulges have a lower effective surface density. 
These results are in good agreement with observations of early--type galaxies 
(see e.g. \citealp{2004MNRAS.355.1155D}).

If compared to nearby merger remnants the range and distribution of S\'ersic indices of the remnants 
are comparable to the observations \citep{2004AJ....128.2098R}. In particular, 
nearby remnants -- similar to simulated remnants -- do not show correlations 
between the surface brightness profiles and other global properties like velocity 
dispersion or isophotal shape. We can therefore conclude that disc galaxies are the likely 
progenitors of the observed remnants. They probably evolve into elliptical galaxies in the 
future. The time-scale for this process, however, depends on the stellar populations of 
the remnants and can in principle be assessed applying evolutionary models for disc 
galaxies (e.g. \citealp{2006MNRAS.tmp...94N}) to N-body models. 

\section*{Acknowledgments}

The authors thank Yago Ascasibar for kindly providing his software to compute the phase-space 
densities. We also thank John Kormendy, Ralf Bender, Roberto Saglia, Alister Graham, 
Fabian Heitsch and Roland 
Jesseit for interestings discussions and valuable comments on the manuscript and Roelof de Jong and 
Barry Rothberg for kindly providing their observational data. Finally we thank the referee Reynier 
Peletier for his valuable comments which improved the manuscript.     
\appendix 

\section{Individual examples} 
\label{EXAMPLES} 

\begin{figure*}
\begin{center}
    \epsfig{file=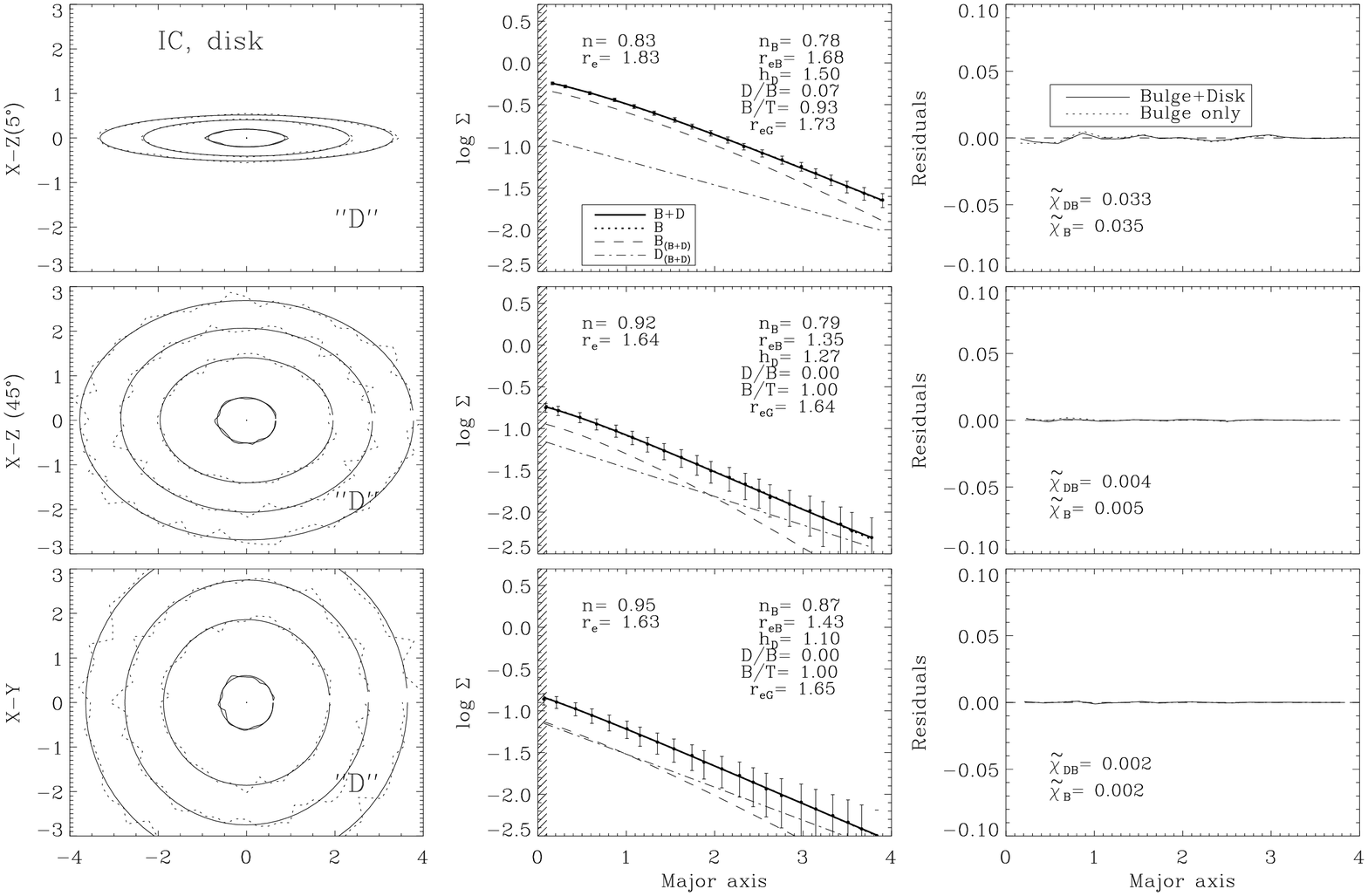,width=0.9\textwidth}
  \caption{Surface density analysis for the initial disc as used for the bulge-less 
mergers. {\it
  Left panel}: Contours of the projected surface density (dotted) with fitted ellipses
  (solid) inclined by $5^o$ (upper plot), $45^o$ (middle plot),  
  and seen face-on (lower plot).
  {\it Middle panel}: Surface density profiles along the major axis of the
  respective projection (dots with error bars). The best fitting
  $DB$-model is shown by the thick solid line. The disc and bulge
  contribution is indicated by the thin dot-dashed and the dashed line,
  respectively. The thick dotted line shows the best fitting S\'ersic-only model, 
  which is in this case close to an exponential. The shaded area indicates the radial range influenced 
by the force softening and has been excluded from the fitted data points. {\it Right panel}:
  Residuals for the two fits and the corresponding reduced chi--squared values. 
This is a one component system with an exponential surface density profile and 
for the face-on projection we can recover the input parameters.  
For $n=1$ the disc scale length $h_D$ is related to the effective radius like 
$r_e = 1.676 h_D$. Observed in projection towards edge-on 
the shape parameter decreases and the size increases by $10\%$. 
 \label{ICN_surf_nacho}} 
\end{center}
\end{figure*}

\begin{figure*}
\begin{center}
    \epsfig{file=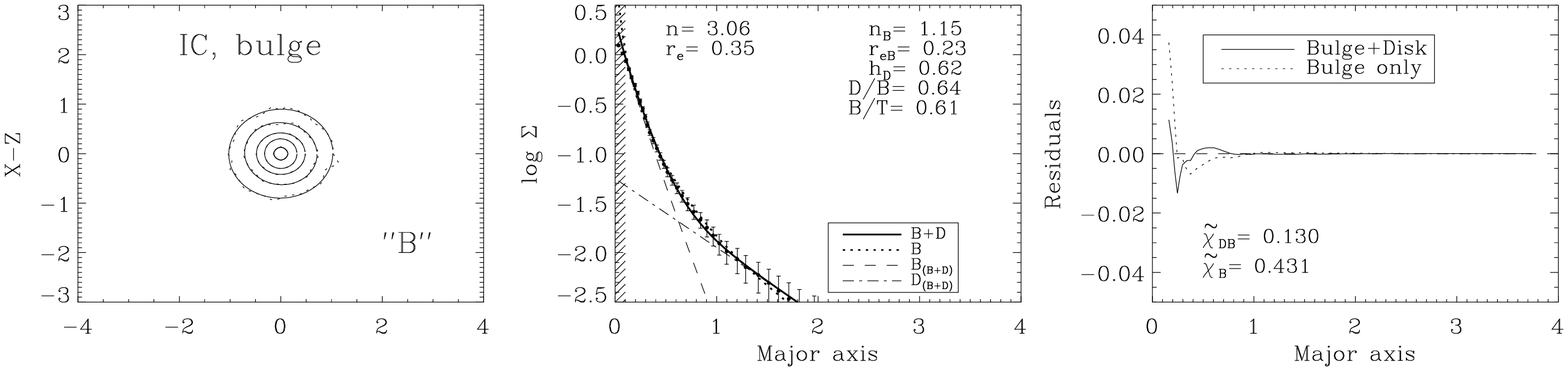,width=0.9\textwidth}
  \caption{Surface density analysis as in Fig. \ref{ICN_surf_nacho} but
 for the initial bulge system only. Note that the projected Hernquist-bulge does 
not perfectly follow an $r^{1/4}$ profile. The theoretically expected value 
for the effective radius, $r_e=0.36$, is similar to the fitted value. The $BD$-fit
does not improve the quality of the fit in this case as the reduced chi--squared 
for the S\'ersic-only fit is already well below unity.}  \label{ICB_surf_nacho}  
\end{center}
\end{figure*}

\begin{figure*}
\begin{center}
    \epsfig{file=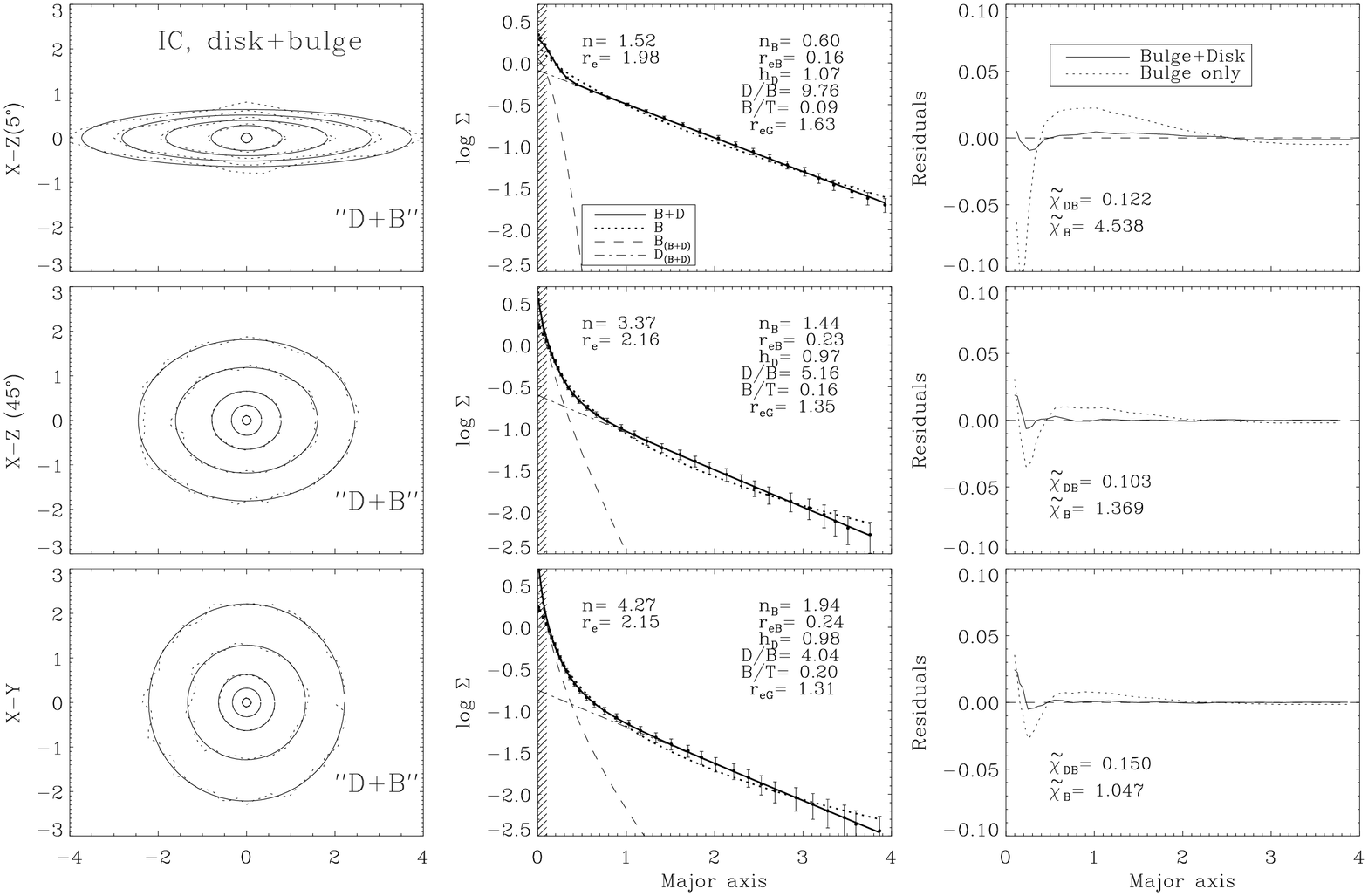,width=0.9\textwidth}
  \caption{Surface density analysis as in Fig. \ref{ICS_surf_nacho} but
for the initial disc+bulge model. For the face-on projection we can nearly 
recover the input parameters. However, the size, mass fraction and curvature 
of the bulge is underestimated if compared to the two individual components (see Figs. 
\ref{ICN_surf_nacho} and \ref{ICB_surf_nacho}). The effect get stronger if the galaxy 
is seen more edge-on.}  
\label{ICS_surf_nacho}   
\end{center}
\end{figure*}

\begin{figure}
\begin{center}
    \epsfig{file=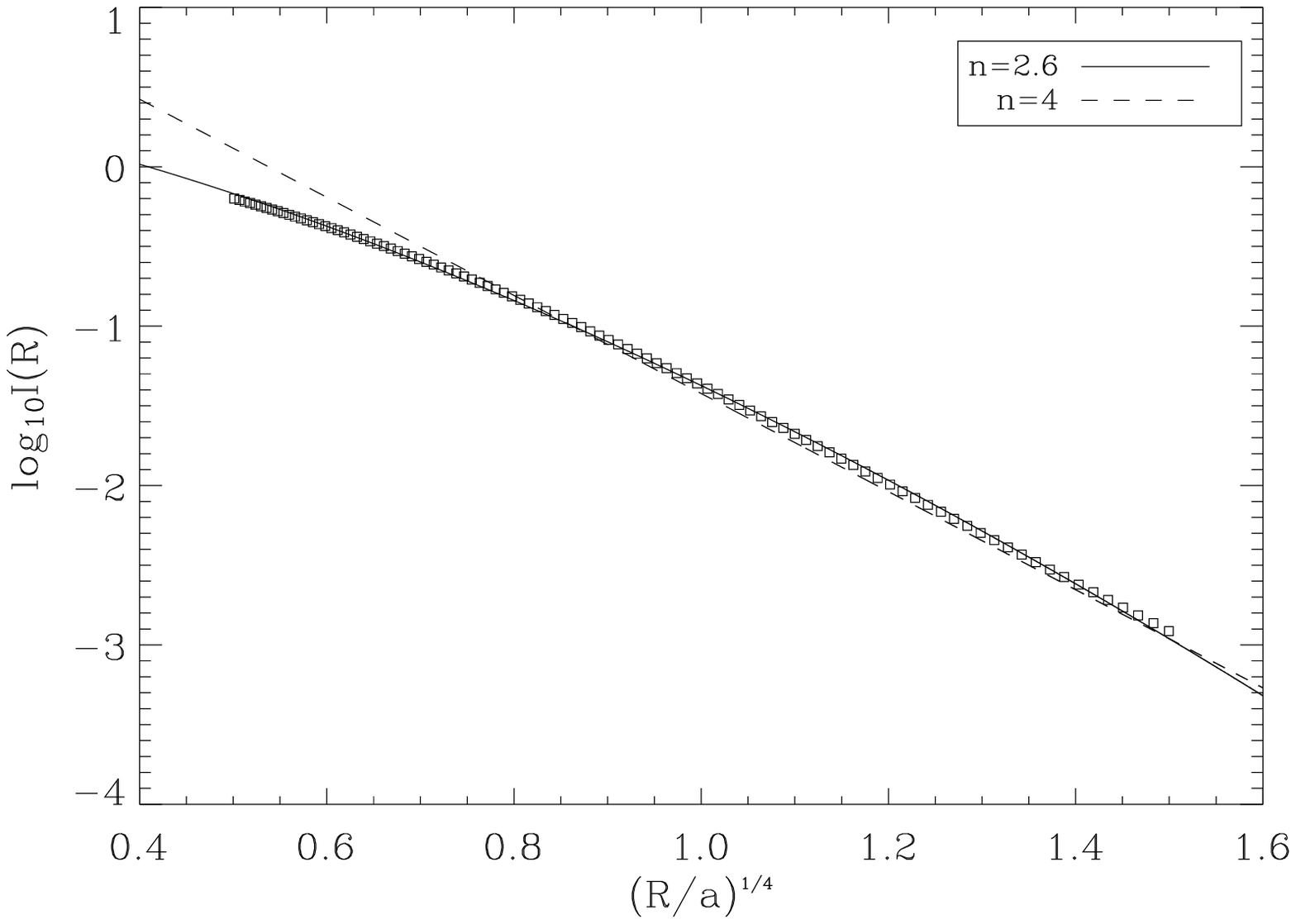,width=0.45\textwidth}
  \caption{Projected \citet{1990ApJ...356..359H} density profile with
    scale length $a$ (open boxes). The best fitting S\'ersic fit (solid
    line) has $n=2.6$ as the projection deviates significantly from a
    \citet{1948AnAp...11..247D} profile with $n=4$ (dashed line) at
    radii $(r/a)^{1/4} < 0.8$. \label{hernquistvsSersic}}       
\end{center}
\end{figure}

\begin{figure}
\begin{center}
    \epsfig{file=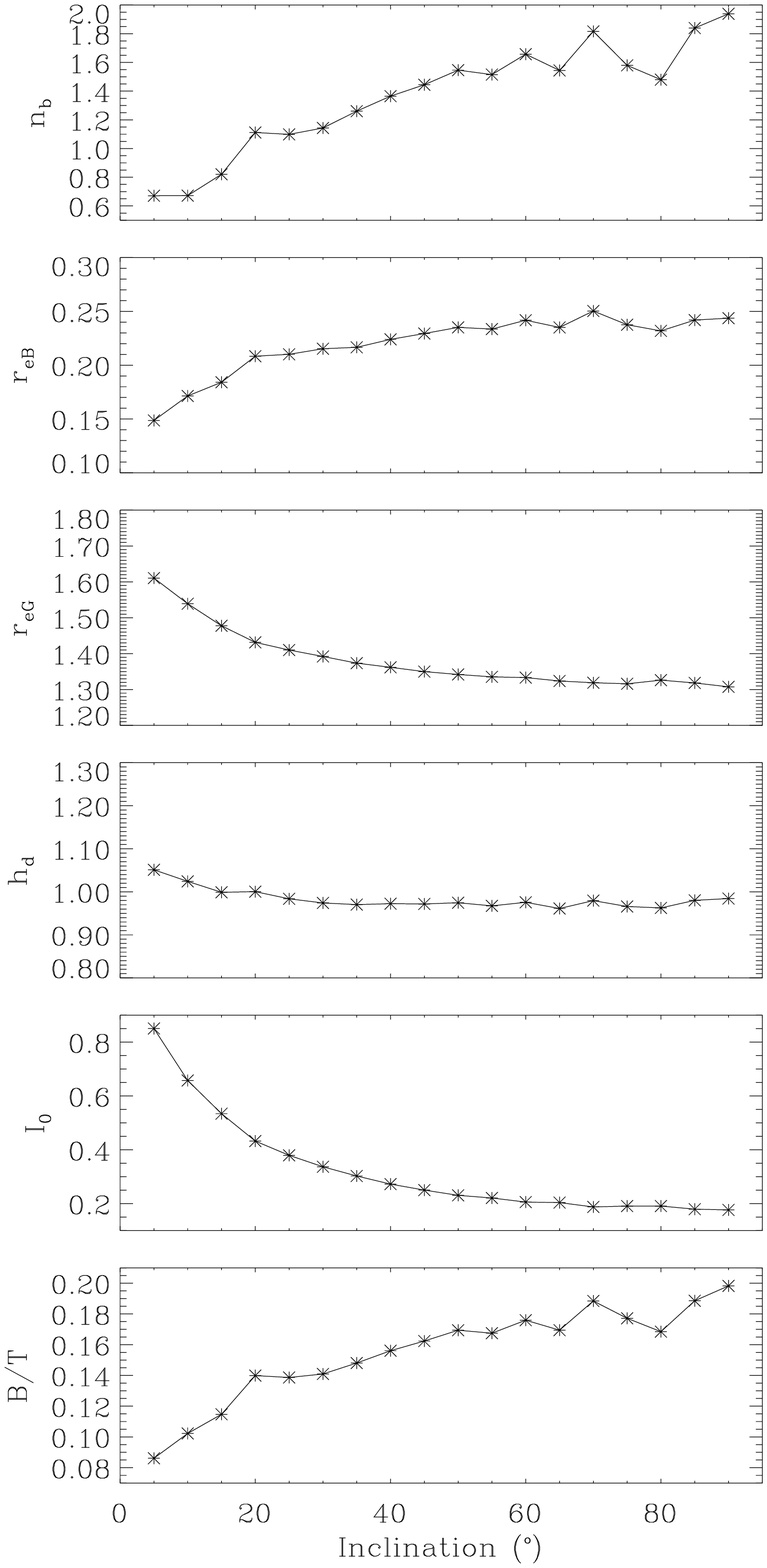,width=0.45\textwidth}
  \caption{Variation of the fitting parameters for the initial condition 
disc+bulge system with inclination angle ($0^o$ is edge-on, $90^o$ is face-on).
The S\'ersic index of the bulge, $n_B$, the size of the bulge, $r_{eB}$, 
and $B/T$ increase from edge-on to face-on, the global size, $r_{eG}$, 
decreases due to an apparently more dominant bulge component. The disc scale length 
$h_d$ does not change with inclination. The central surface density $I_0$ of the disc 
is highest for the edge-on case and decreases strongly towards face-on.   
\label{ang_var_nacho}}
\end{center}
\end{figure}
To demonstrate the quality and the limitations of the surface density
analysis we show and discuss the surface density fits of the initial 
condition disc and disc+bulge model as well as of one merger remnant of 
every mass ratio, respectively. 
\subsection{Initial conditions} 
\label{INITIAL}
Fig. \ref{ICN_surf_nacho} shows the real isophotes and the best fitting
ellipses, the surface density analysis and the residuals for the
initial condition disc-only model as seen tilted by $5^o$, $45^o$ and
face-on. The parameters for the S\'ersic--only and the bulge+disc fits
are given in the middle panel. All three projections of the system are
classified as one component systems with close to exponential density profiles. 
The analysis program was able to recover the initial condition parameters 
almost perfectly if the disc is seen face-on (last row in Fig. \ref{ICN_surf_nacho}). 
If the galaxy is seen more edge-on the shape parameter decreases and the size 
increases by more than 10\%. Note that for a S\'ersic--profile with 
$n=1$ (exponential profile) the scale length $h_D$ is related to the 
projected effective radius like 
$r_e = 1.676 h_D$. 

In Fig. \ref{ICB_surf_nacho} we show the analysis for the bulge component separately.  
The scale length $a_H$ of the Hernquist-spheroid  is related to its effective radius, 
$r_{eH}$, as $r_{eH}= a_H (1+\sqrt{a_H})/1.33$. For the model used here with a scale 
length of $a_H=a_b=0.2$ this would result in $r_{eH} =r_{eB} = 0.36$. This value can be recovered 
by the fit.   We note here that the projected
Hernquist density distribution follows a de Vaucouleurs like $r^{1/4}$
profile only at radii larger than half its scale length. If analysed over the full radial range
the best fitting S\'ersic-index for the projected
\citet{1990ApJ...356..359H} density distribution would be $n=2.6$   
(see Fig. \ref{hernquistvsSersic}) which is lower than the value of our bulge, $n=3.06$, 
(see Fig. \ref{ICB_surf_nacho}) as we exclude the more flattened innermost parts from the fit. 

The analysis for the composite disc+bulge model is shown in Fig. 
\ref{ICS_surf_nacho}. The fitted scale length of the disc seems in good 
agreement with the initial parameters (Section \ref{PARAMETERS}), 
independent of projection. The best agreement with the theoretically expected 
values is achieved for the face-on projection. The disc--to--bulge ratio
$D/B = 4.04$  is 25\% larger than the dynamical input value of $M_d/M_b = 3$. The shape 
parameter of the bulge, $n_B = 1.94$, is smaller than for the isolated bulge.  
The maximum projected size of the bulge is equal to the isolated bulge model. 
If projected nearly edge-on the bulge--to--total 
ratio drops below $B/T = 0.1$ and the system appears to be disc dominated except from the very inner
parts. The apparent size and S\'ersic-index of the
bulge in the edge-on projection are significantly smaller. Part of the outer bulge material 
is counted as disc material to guarantee an exponential projected profile -- which 
does not represent the intrinsic disc distribution for this projection 
(see Fig. \ref{ICN_surf_nacho}). The bulge can then very well be fitted with an exponential 
as only the inner part of the bulge is accessible. In Fig. \ref{ang_var_nacho}
 we show how the fitting parameters change with disc inclination. The shape index of the bulge 
changes almost linearly from $n_B \approx 0.6$ in the edge-on projection to 
$n_B \approx 2$ in the face-on projection. The bulge is also smaller if seen edge-on. The system as a 
whole appears larger in the edge-on projection. The global effective radius of the initial bulge+disc
system has been calculated using Eqn. \ref{re_global}  and is in the range
of $ 1.3 < r_{eG} < 1.6$. The values are lower than for the pure disc system 
(Fig. \ref{ICN_surf_nacho}) due to the additional centrally concentrated bulge component.
The disc scale length, $h_d$, does not change with projection, in contrast to the 
disc only model (Fig. \ref{ICN_surf_nacho}). 
The central surface density of the disc, however, changes strongly with 
inclination, especially at inclinations smaller than $40^o$ (towards edge-on). This effect 
has already been investigated with a simple model (see e.g. \citealp{2001MNRAS.326..543G}).  
The bulge-to-total ratio is in the range of $0.09 < B/T < 0.2$ and is increasing from 
edge-on to face-on. The increase is strongly correlated with the increase of $n_B$.  
These results indicate that observed disc--to--bulge ratios of early-type disk galaxies might 
be systematically overestimated and S\'ersic-indices of bulges might be systematically underestimated if 
the galaxies are inclined with respect to the face-on projection. 
This effect might be partly responsible for part of the scatter in the observed $B/T-n$ relation 
\citep{1995MNRAS.275..874A} and we propose to investigate    
projection effects of observed disc galaxies in more detail. It also seems difficult
to reconstruct the underlying mass distribution of the disc and the bulge as the low density 
part of the bulge is masked by the disc. In this way 20\% - 35\% of the bulge material (depending 
on the projection) could remain undetected in observed disc galaxies.  The above results are stable
and do not change significantly if we vary the fitting parameters like weights, inner boundaries, 
starting values etc. However, the details might depend 
very well on our choice of the initial conditions. The Hernquist spheroid is a good but not 
the optimal model for real bulges and more sophisticated bulge models that reproduce 
the observed properties might be useful for further investigations. \\    
\begin{figure*}
\begin{center}
    \epsfig{file=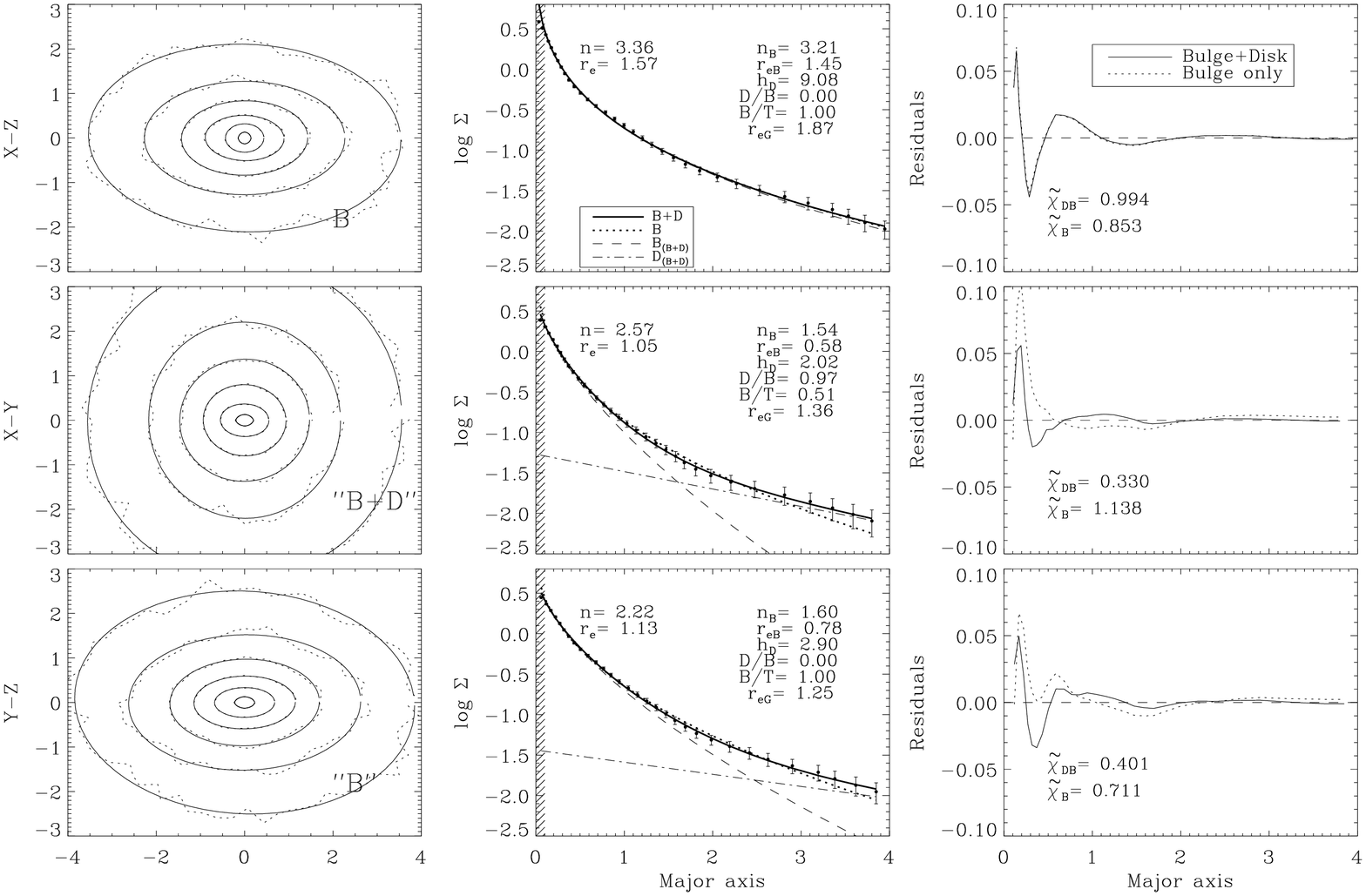,width=0.9\textwidth}
  \caption{Surface density analysis as in Fig. \ref{ICS_surf_nacho} but
 for a 1:1 merger remnant projected along its three principal
 axes. Based on the selection criteria described in Section
 \ref{CLASSIFICATION} the remnant would be classified as a pure bulge
 system in the first and third projections and as a bulge+disc system in the second projection.} 
  \label{11_surf_nacho}  
\end{center}
\end{figure*}
\begin{figure*}
\begin{center}
    \epsfig{file=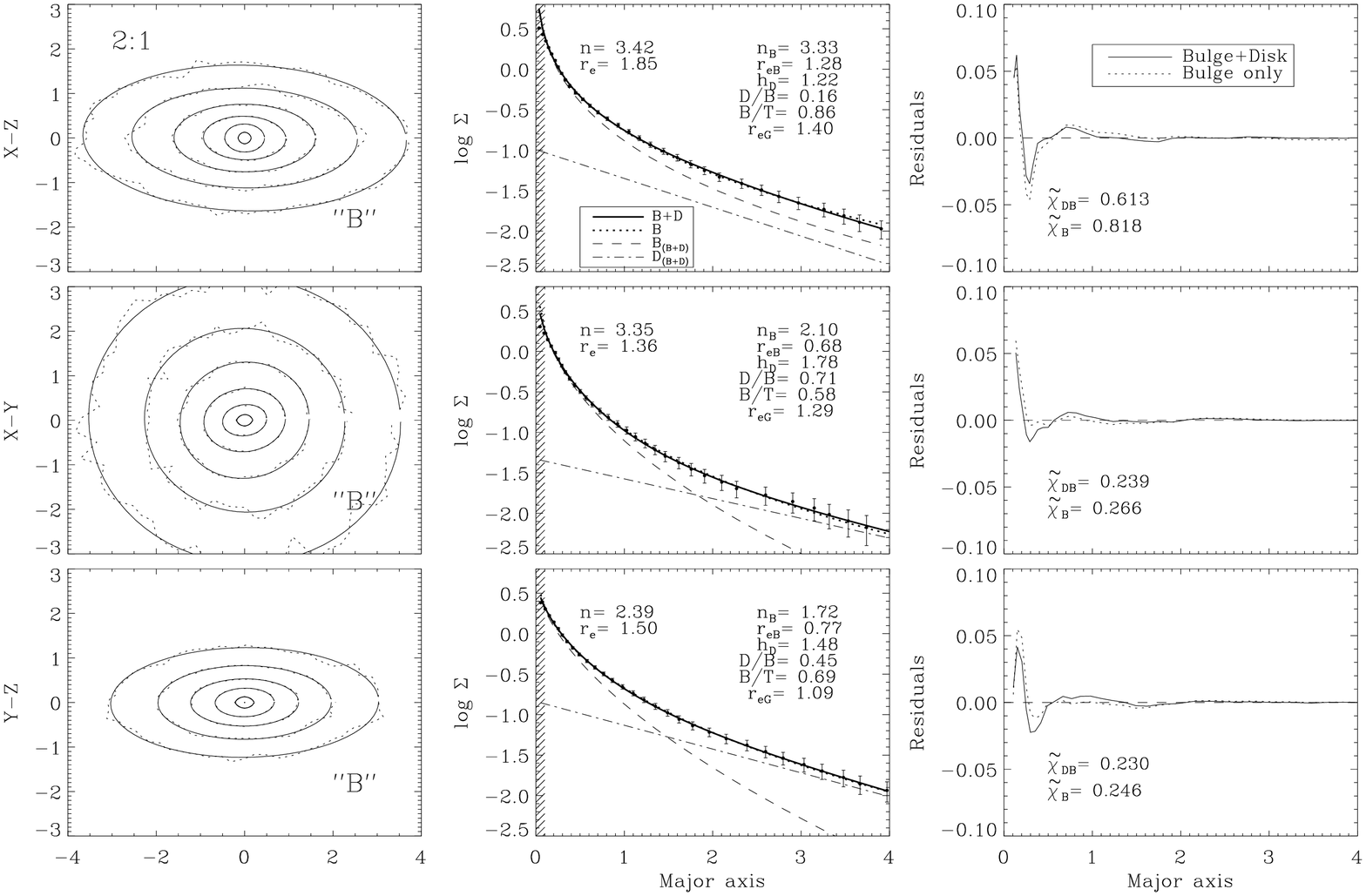,width=0.9\textwidth}
  \caption{Same as Fig. \ref{11_surf_nacho} but for a 2:1 remnant. If
    observed, the remnant would be classified as a bulge in all three projections.}
  \label{21_surf_nacho}   
\end{center}
\end{figure*}
\begin{figure*}
\begin{center}
    \epsfig{file=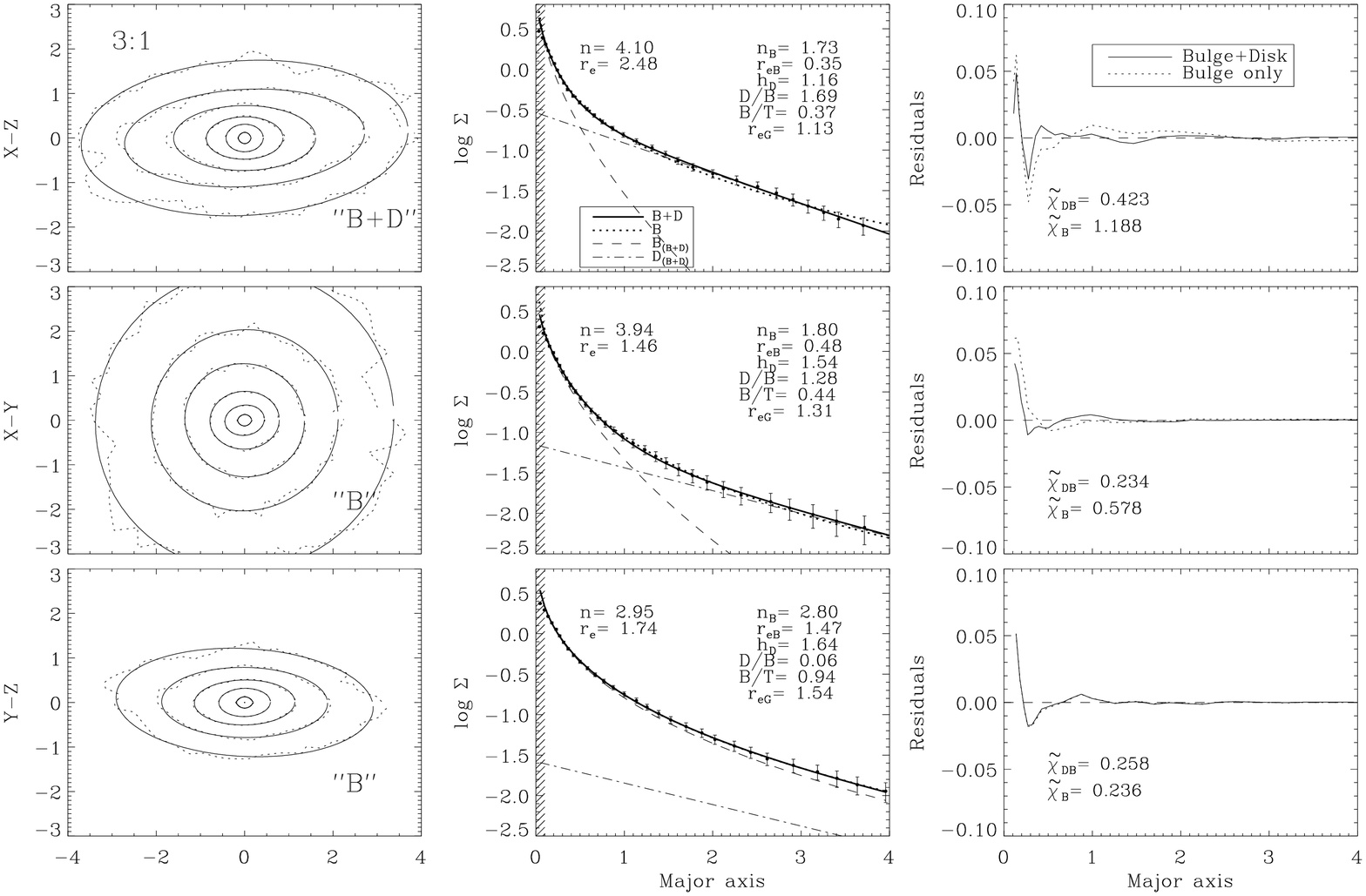,width=0.9\textwidth}
  \caption{Same as Fig. \ref{11_surf_nacho} but for a 3:1 remnant. If
  observed, the remnant would be classified as a disk+bulge system in the
  projection along the intermediate axis (first row) and as a pure bulge system
 in the other two projections.}
  \label{31_surf_nacho}   
\end{center}
\end{figure*}
\begin{figure*}
\begin{center}
    \epsfig{file=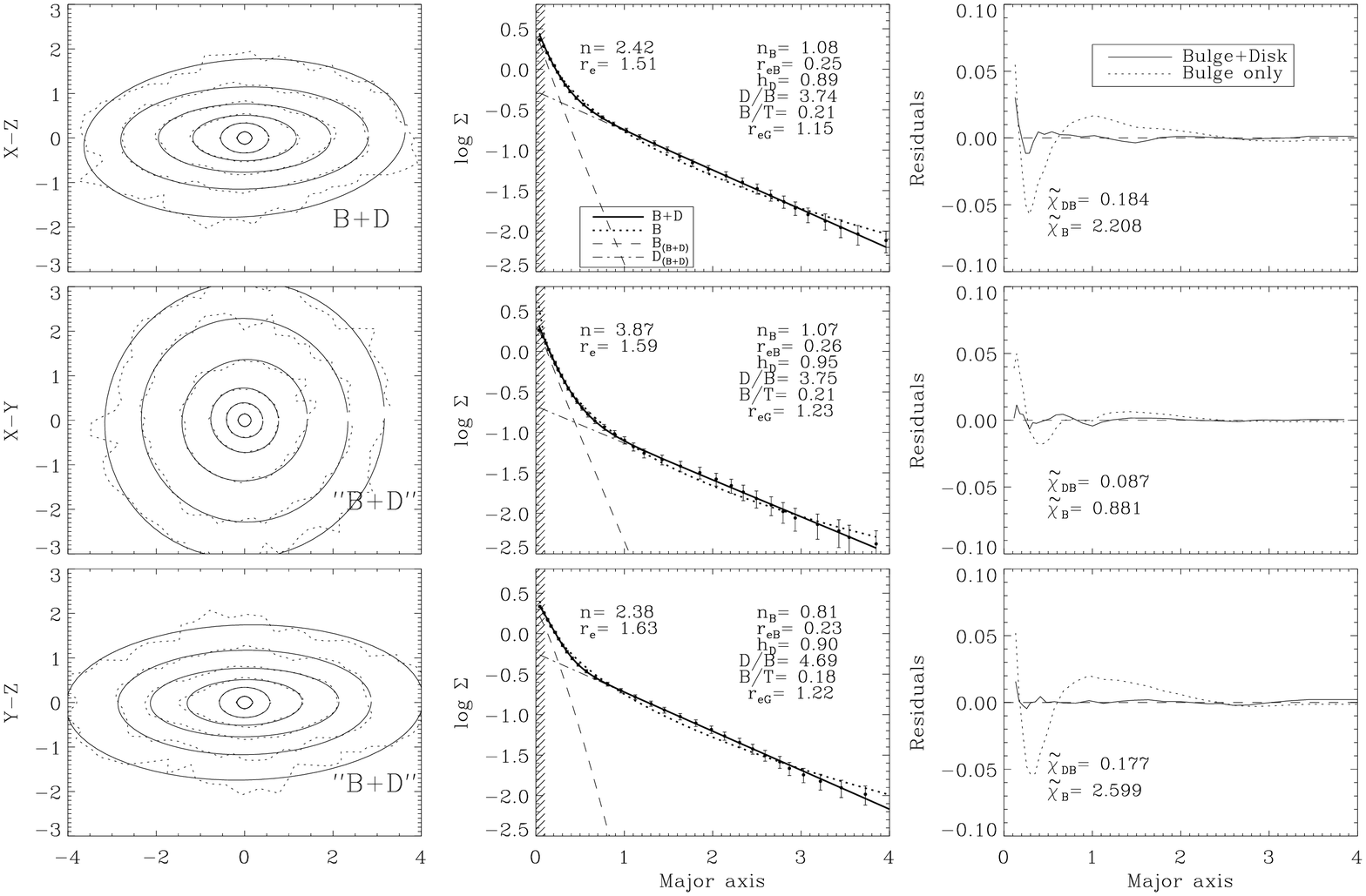,width=0.9\textwidth} 
    \caption{Same as Fig. \ref{11_surf_nacho} but for a 4:1
      remnant. If observed, the remnant would be classified as a
      disc+bulge system in all three  projections. \label{41_surf_nacho}}  
\end{center}
\end{figure*}
\begin{figure*}
\begin{center}
    \epsfig{file=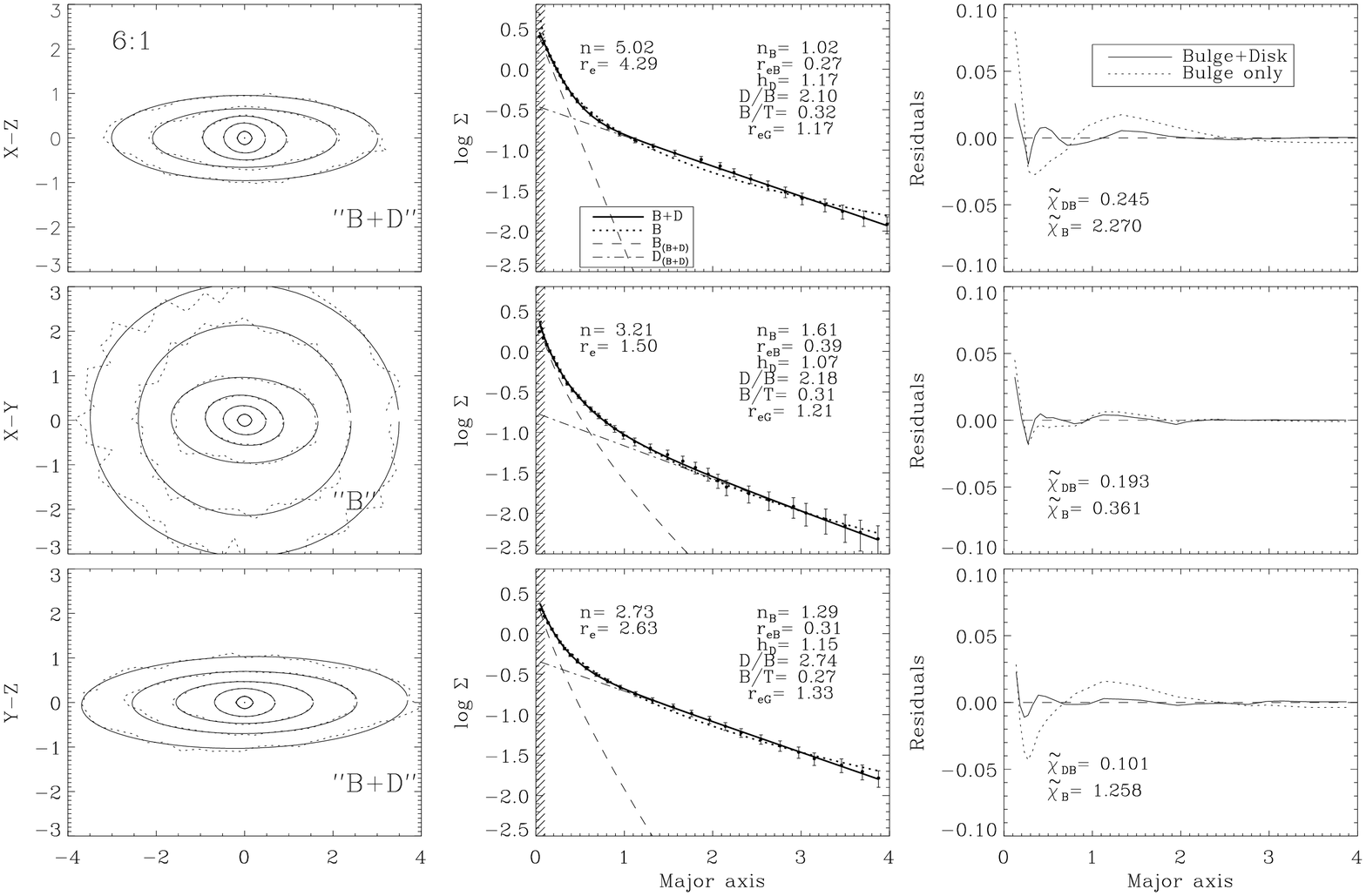,width=0.9\textwidth} 
    \caption{Same as Fig. \ref{11_surf_nacho} but for a 6:1
      remnant. If observed, the remnant would be classified as a
      disc+bulge system in the two edge-on  projections.}
      \label{61_surf_nacho}  
\end{center}
\end{figure*}
\begin{figure*}
\begin{center}
    \epsfig{file=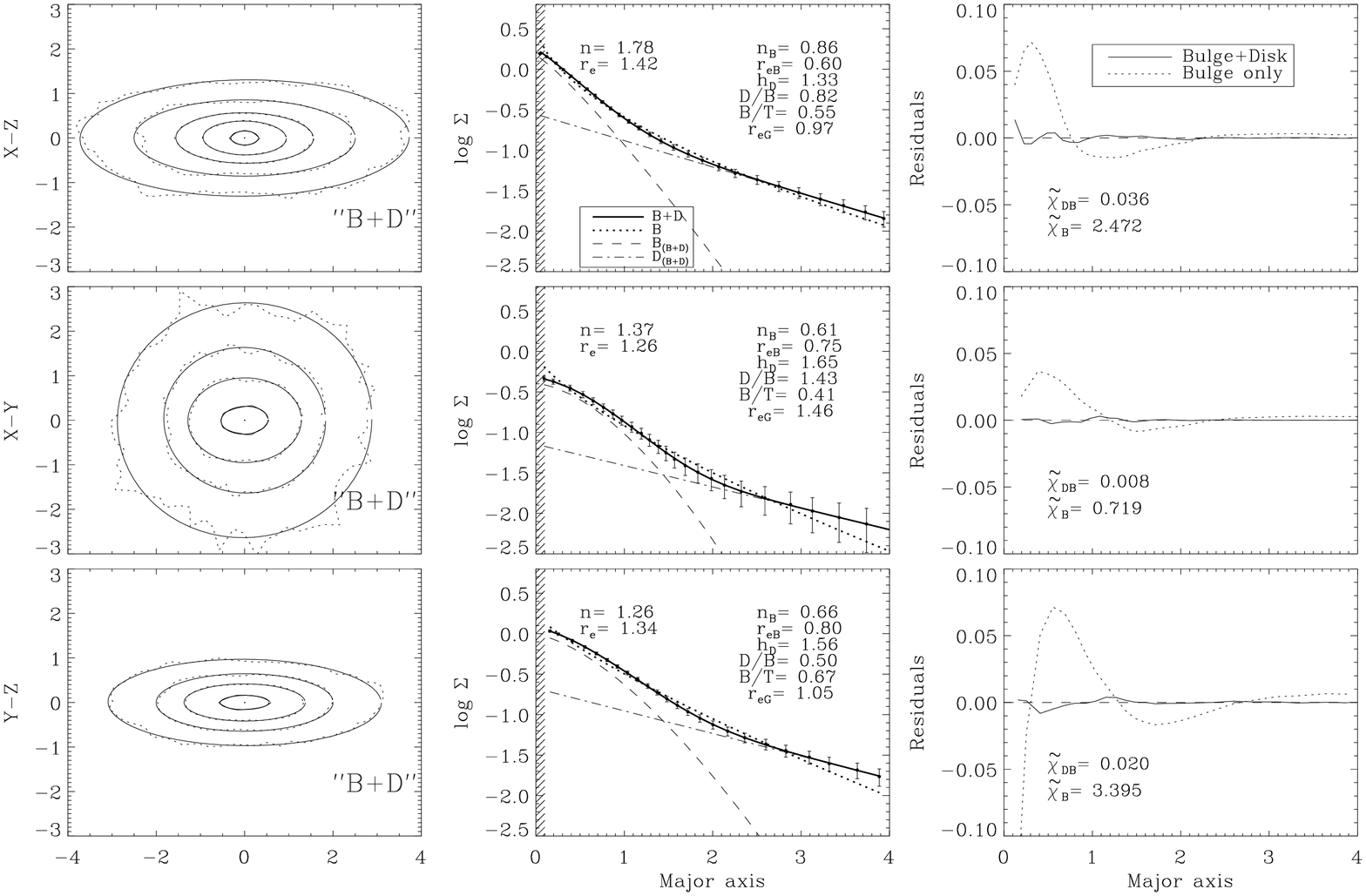,width=0.9\textwidth} 
    \caption{Same as Fig. \ref{11_surf_nacho} but for a 1:1 
      remnant without a bulge in the progenitor disc. 
 If observed, the remnant would be classified as a
      disc+bulge system in all projections. The system is significantly less concentrated
    than the 1:1 remnant with bulge (Fig. \ref{11_surf_nacho}).}
      \label{11N_surf_nacho}  
\end{center}
\end{figure*}
\begin{figure*}
\begin{center}
    \epsfig{file=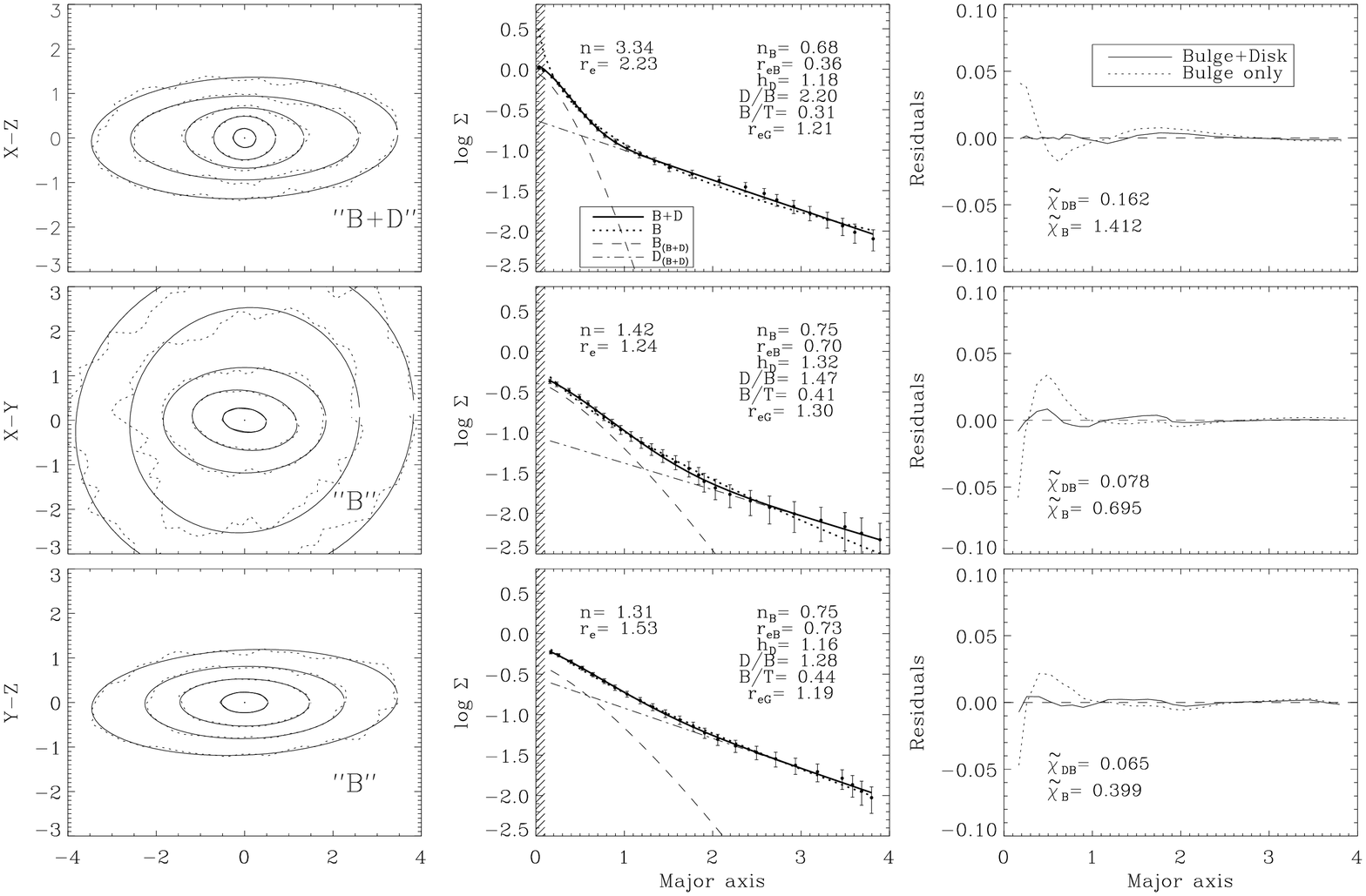,width=0.9\textwidth} 
    \caption{Same as Fig. \ref{11_surf_nacho} but for a bulge-less 3:1
      remnant. If observed, the remnant would be classified as a disc+bulge in the first and 
      as a bulge in the second and third projection. The remnant is prolate at the center 
and appears more concentrated if seen along the prolate structure (first row).}
      \label{31N_surf_nacho}  
\end{center}
\end{figure*}

\subsection{Individual merger remnants}
In Figs. \ref{11_surf_nacho} to \ref{61_surf_nacho} we show individual 
examples of the surface density analysis of disc+bulge merger remnants with mass 
ratios of 1:1, 2:1, 3:1, 4:1, and 6:1, respectively. A 1:1 and 3:1 remnant of
mergers of two pure disc galaxies are shown in Figs.  \ref{11N_surf_nacho} and 
\ref{31N_surf_nacho}. All remnants are projected along the
three principal axes of their moment-of-inertia tensor calculated form the $40\%$
most tightly bound particles. The first projection in
Fig. \ref{11_surf_nacho} represents a typical system that is classified
as a one component bulge system. The freedom of adding an 
exponential component does
not increase the quality of the fit. The second projection is
classified as a disc+bulge system based on its bulge--to--total ratio
and the better quality of the fit. The third projection would be classified as
a one-component bulge system as the reduced chi--squared of the fit is below unity and $P < 0.32$ 
(Eqn. \ref{prob}). To test for the effect of resolution we have re-simulated this remnant 
with a factor of 2 and 4 more particles and did not find a change in the properties exceeding 
the errors of the fitting parameters (Section \ref{STABILITY}).  

All three projections of the 2:1 remnant shown in Fig. \ref{21_surf_nacho} can be fitted by a
pure bulge model with sufficient accuracy. The first projection of the 3:1
remnant in Fig. \ref{31_surf_nacho} is classified as a disc+bulge
system whereas the other two projections appear to be pure bulge
systems. An example for a system with strong indications for an outer exponential component 
is the 4:1 merger remnant shown in Fig. \ref{41_surf_nacho}. The first and third 
projections (``edge-on'') clearly show two component systems. For the face-on projection  
both fits would result in a reduced  chi--squared below one. However, only for the $DB$ fit 
$P < 0.32$ and the system is classified as a $BD$ system. The 6:1 remnants shown in Fig. 
\ref{61_surf_nacho} shows prominent exponential components in the "edge-on" projections. 
However, if seen face-on it would still be classified as a single bulge.
Fig. \ref{11N_surf_nacho} shows the surface density profile of a bulge-less 1:1 merger remnant. 
Due to the absence of the bulge the profile is significantly flatter at small radii 
than the corresponding remnant with bulge (Fig. \ref{11_surf_nacho}).  
In all the projections the remnant does not have the characteristics of a single component system. 
The inner component is better represented by S\'ersic function with a shape parameter $n_B < 1$, 
even close to a Gaussian with $n=0.5$. A 3:1 remnant without bulge is shown in Fig. \ref{31N_surf_nacho}.
In the first projection the remnant is classified as a two component system with a very 
shallow inner component. The two remaining projections appear like one component systems, however, 
with shape indices $n < 2$. This remnant has very prolate structure at the center which is typical for 
remnants without bulges. They appear more concentrated if projected along this structure (first rows in 
Fig. \ref{11N_surf_nacho} and Fig.\ref{31N_surf_nacho}).

Already this first analysis clarifies that different projections of
one merger remnant can lead to different fitting parameters. 
The ranges covered in $B/T$ and S\'ersic-indices are large and, in most
cases, the classification of the merger remnant changes depending on the projection.  
It is therefore not possible to make any reliable statement based on
individual projections of merger remnants and only a statistical
analysis, as presented in Section \ref{STATISTIC}, of the whole 
homogenous sample of simulations will allow us to draw any global
conclusions.   

\subsection{Stability of the analysis} 
\label{STABILITY}
\begin{figure}
\begin{center}
    \epsfig{file=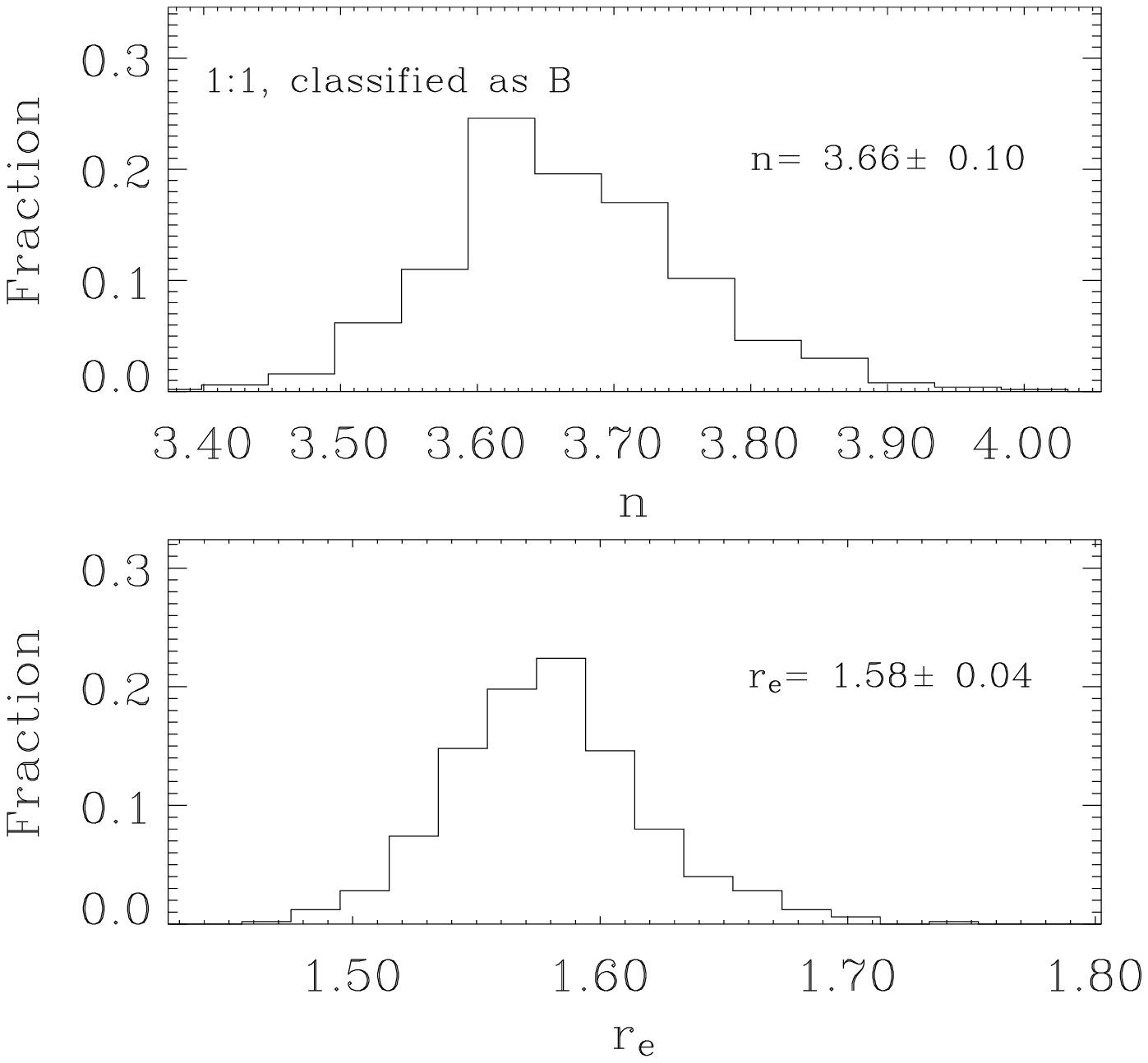,width=0.5\textwidth}
  \caption{Typical distribution of the S\'ersic-index (upper panel) and the
  effective radius (lower panel) for a fixed projection of a 1:1
  merger remnant classified as a bulge for 500 independent
  realizations using the boots-trapping method. The distributions are
  narrow and the errors are well below 5\%.}\label{11_stat_nacho}   
\end{center}
\end{figure}
\begin{figure}
\begin{center}
    \epsfig{file=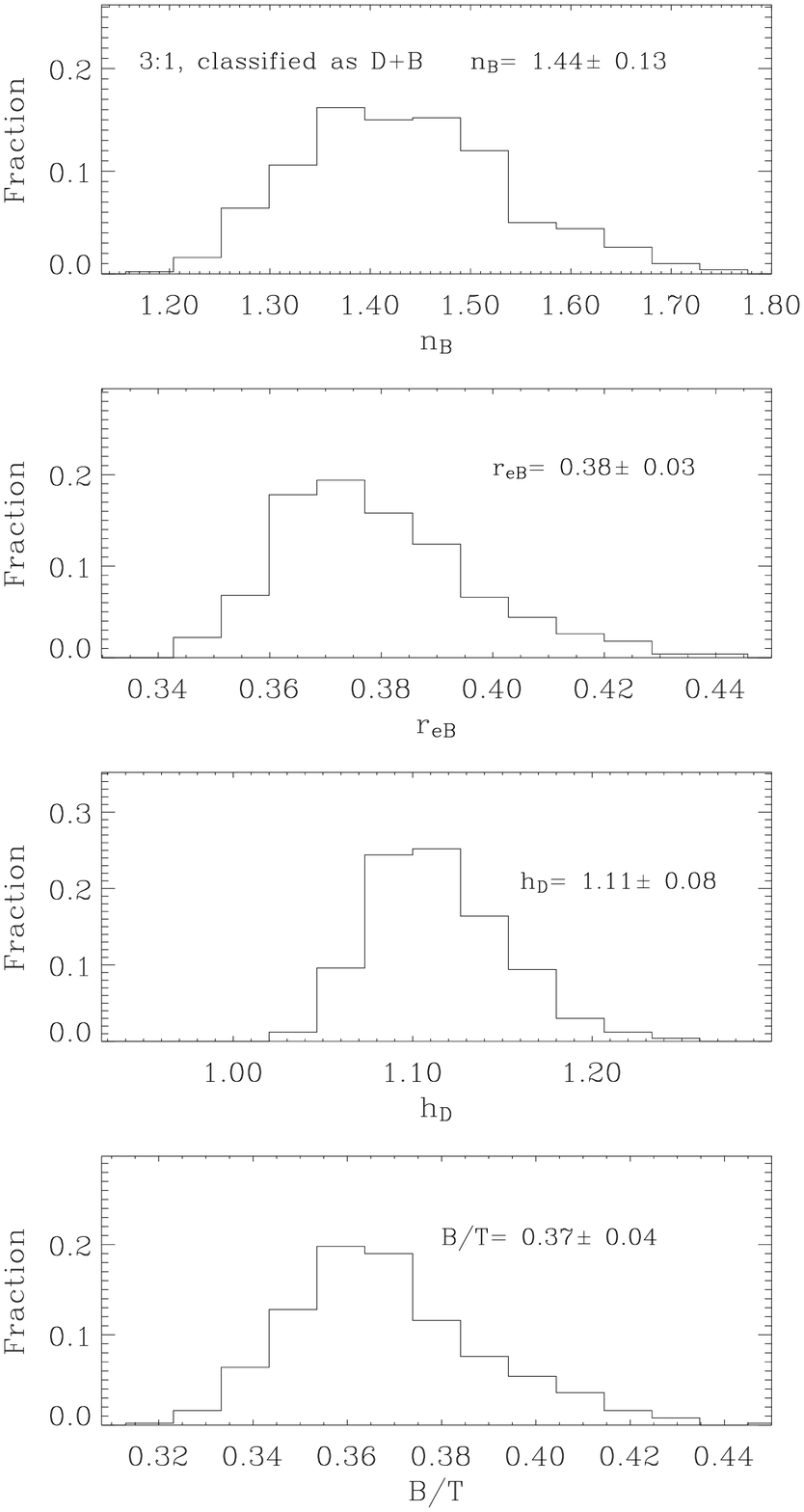,width=0.5\textwidth}
  \caption{Boots-trapping analysis for 500 independent realizations
  (as in Fig. \ref{11_stat_nacho}) of a typical 3:1 merger remnant
  classified as a disc+bulge system seen from a fixed projection. The
  errors for the S\'ersic-index of the bulge, the effective radius of
  the bulge, the disc scale length and the bulge--to--total ratio do not
  exceed the 10\% level.}   \label{31_stat_nacho}  
\end{center}
\end{figure}
We used the boots-trapping method
\citep{1994ApJ...427..165H,2003ApJ...597..893N} to estimate the typical errors
of the derived fit parameters. For this analysis we created 500  
re-sampled realizations of one projection of the original N-particle
distribution by selecting N particles from the remnant with
replacement. For every realization some particles might appear several
times and some might not appear at all. Thereafter we performed a full
surface density analysis for all 500 realizations and computed the
mean values and the standard deviations of the resulting fit
parameters. We have selected two typical merger remnants as test
cases: a projected 1:1 remnant that was classified as a
were described primarily by the exponential component. To avoid these cases,
the algorithm checks whether the central intensity of the fitted exponential
component is larger than the central intensity of the fitted S\'ersic bulge system
and a projected 3:1 remnant classified as a disc+bulge system.  

Fig. \ref{11_stat_nacho} shows the distribution of values for the 
S\'ersic-index and the effective radius for the 1:1 test case which 
was classified as a bulge.  In both cases the distribution is very
narrow and the error is below 5\%. The distribution for the four
fitted values --- S\'ersic-index of the bulge, effective radius of the
bulge, scale length of the disc and bulge--to--total ratio --- of the
3:1 remnant classified as DB is shown in
Fig. \ref{31_stat_nacho}. The errors are larger than for the 1:1
remnants  but still only of the order of 10\%.  The most important
result of this analysis  is that the variation in properties for
different projections as shown in the previous section is real and not
dominated by errors. The errors given here are typical and can be used 
for all following plots.

\bibliographystyle{mn2e}
\bibliography{references}

\begin{thebibliography}{}

\bibitem[\protect\citeauthoryear{{Aceves} \& {Vel{\' a}zquez}}{{Aceves} \&
  {Vel{\' a}zquez}}{2005}]{2005MNRAS.360L..50A}
{Aceves} H.,  {Vel{\' a}zquez} H.,  2005, \mnras, 360, L50

\bibitem[\protect\citeauthoryear{{Aguerri} \& {Trujillo}}{{Aguerri} \&
  {Trujillo}}{2002}]{2002MNRAS.333..633A}
{Aguerri} J.~A.~L.,  {Trujillo} I.,  2002, \mnras, 333, 633

\bibitem[\protect\citeauthoryear{{Andredakis}, {Peletier} \&
  {Balcells}}{{Andredakis} et~al.}{1995}]{1995MNRAS.275..874A}
{Andredakis} Y.~C.,  {Peletier} R.~F.,    {Balcells} M.,  1995, \mnras, 275,
  874

\bibitem[\protect\citeauthoryear{{Ascasibar} \& {Binney}}{{Ascasibar} \&
  {Binney}}{2005}]{2005MNRAS.356..872A}
{Ascasibar} Y.,  {Binney} J.,  2005, \mnras, 356, 872

\bibitem[\protect\citeauthoryear{{Barnes}}{{Barnes}}{1992}]{1992ApJ...393..484%
B}
{Barnes} J.~E.,  1992, \apj, 393, 484

\bibitem[\protect\citeauthoryear{{Barnes}}{{Barnes}}{1998}]{1998giis.conf..275%
B}
{Barnes} J.~E.,  1998, in Saas-Fee Advanced Course 26: Galaxies: Interactions
  and Induced Star Formation {Dynamics of Galaxy Interactions}.
pp 275--+

\bibitem[\protect\citeauthoryear{{Barnes}}{{Barnes}}{2002}]{2002MNRAS.333..481%
B}
{Barnes} J.~E.,  2002, \mnras, 333, 481

\bibitem[\protect\citeauthoryear{{Bekki}}{{Bekki}}{1998}]{1998ApJ...502L.133B}
{Bekki} K.,  1998, \apjl, 502, L133+

\bibitem[\protect\citeauthoryear{{Bell}, {Naab}, {McIntosh}, {Somerville},
  {Caldwell}, {Barden}, {Wolf}, {Rix}, {Beckwith}, {Borch}, {Haeussler},
  {Heymans}, {Jahnke}, {Jogee}, {Meisenheimer}, {Peng}, {Sanchez} \&
  {Wisotzki}}{{Bell} et~al.}{2005}]{2005astro.ph..6425B}
{Bell} E.~F.,  {Naab} T.,  {McIntosh} D.~H.,  {Somerville} R.~S.,  {Caldwell}
  J.~A.~R.,  {Barden} M.,  {Wolf} C.,  {Rix} H.-W.,  {Beckwith} S.~V.~W.,
  {Borch} A.,  {Haeussler} B.,  {Heymans} C.,  {Jahnke} K.,  {Jogee} S.,
  {Meisenheimer} K.,  {Peng} C.~Y.,  {Sanchez} S.~F.,    {Wisotzki} L.,  2005,
  ArXiv Astrophysics e-prints

\bibitem[\protect\citeauthoryear{{Binggeli} \& {Jerjen}}{{Binggeli} \&
  {Jerjen}}{1998}]{1998A&A...333...17B}
{Binggeli} B.,  {Jerjen} H.,  1998, \aap, 333, 17

\bibitem[\protect\citeauthoryear{{Binney}}{{Binney}}{2005}]{2005astro.ph..4387%
B}
{Binney} J.,  2005, ArXiv Astrophysics e-prints

\bibitem[\protect\citeauthoryear{{Bournaud}, {Combes} \& {Jog}}{{Bournaud}
  et~al.}{2004}]{2004A&A...418L..27B}
{Bournaud} F.,  {Combes} F.,    {Jog} C.~J.,  2004, \aap, 418, L27

\bibitem[\protect\citeauthoryear{{Bournaud}, {Jog} \& {Combes}}{{Bournaud}
  et~al.}{2005}]{2005A&A...437...69B}
{Bournaud} F.,  {Jog} C.~J.,    {Combes} F.,  2005, \aap, 437, 69

\bibitem[\protect\citeauthoryear{{Burkert}}{{Burkert}}{1993}]{1993A&A...278...%
23B}
{Burkert} A.,  1993, \aap, 278, 23

\bibitem[\protect\citeauthoryear{{Burkert} \& {Naab}}{{Burkert} \&
  {Naab}}{2005}]{2005astro.ph..4595B}
{Burkert} A.,  {Naab} T.,  2005, ArXiv Astrophysics e-prints

\bibitem[\protect\citeauthoryear{{Caon}, {Capaccioli} \& {D'Onofrio}}{{Caon}
  et~al.}{1993}]{1993MNRAS.265.1013C}
{Caon} N.,  {Capaccioli} M.,    {D'Onofrio} M.,  1993, \mnras, 265, 1013

\bibitem[\protect\citeauthoryear{{Caon}, {Capaccioli} \& {D'Onofrio}}{{Caon}
  et~al.}{1994}]{1994A&AS..106..199C}
{Caon} N.,  {Capaccioli} M.,    {D'Onofrio} M.,  1994, \aaps, 106, 199

\bibitem[\protect\citeauthoryear{{Caon}, {Capaccioli} \& {Rampazzo}}{{Caon}
  et~al.}{1990}]{1990A&AS...86..429C}
{Caon} N.,  {Capaccioli} M.,    {Rampazzo} R.,  1990, \aaps, 86, 429

\bibitem[\protect\citeauthoryear{{Capaccioli}}{{Capaccioli}}{1987}]{1987IAUS..%
127...47C}
{Capaccioli} M.,  1987, in IAU Symp. 127: Structure and Dynamics of Elliptical
  Galaxies {Distribution of light - Outer regions}.
pp 47--60

\bibitem[\protect\citeauthoryear{{Carlberg}}{{Carlberg}}{1986}]{1986ApJ...310.%
.593C}
{Carlberg} R.~G.,  1986, \apj, 310, 593

\bibitem[\protect\citeauthoryear{{Cox}, {Jonsson}, {Primack} \&
  {Somerville}}{{Cox} et~al.}{2005}]{2005astro.ph..3201C}
{Cox} T.~J.,  {Jonsson} P.,  {Primack} J.~R.,    {Somerville} R.~S.,  2005,
  ArXiv Astrophysics e-prints

\bibitem[\protect\citeauthoryear{{Davies}, {Phillipps}, {Cawson}, {Disney} \&
  {Kibblewhite}}{{Davies} et~al.}{1988}]{1988MNRAS.232..239D}
{Davies} J.~I.,  {Phillipps} S.,  {Cawson} M.~G.~M.,  {Disney} M.~J.,
  {Kibblewhite} E.~J.,  1988, \mnras, 232, 239

\bibitem[\protect\citeauthoryear{{de Jong}, {Simard}, {Davies}, {Saglia},
  {Burstein}, {Colless}, {McMahan} \& {Wegner}}{{de Jong}
  et~al.}{2004}]{2004MNRAS.355.1155D}
{de Jong} R.~S.,  {Simard} L.,  {Davies} R.~L.,  {Saglia} R.~P.,  {Burstein}
  D.,  {Colless} M.,  {McMahan} R.,    {Wegner} G.,  2004, \mnras, 355, 1155

\bibitem[\protect\citeauthoryear{{de Vaucouleurs}}{{de
  Vaucouleurs}}{1948}]{1948AnAp...11..247D}
{de Vaucouleurs} G.,  1948, Annales d'Astrophysique, 11, 247

\bibitem[\protect\citeauthoryear{{Freeman}}{{Freeman}}{1970}]{1970ApJ...160..8%
11F}
{Freeman} K.~C.,  1970, \apj, 160, 811

\bibitem[\protect\citeauthoryear{{Gonz{\' a}lez-Garc{\'{\i}}a} \&
  {Balcells}}{{Gonz{\' a}lez-Garc{\'{\i}}a} \&
  {Balcells}}{2005}]{2005MNRAS.357..753G}
{Gonz{\' a}lez-Garc{\'{\i}}a} A.~C.,  {Balcells} M.,  2005, \mnras, 357, 753

\bibitem[\protect\citeauthoryear{{Gonz{\' a}lez-Garc{\'{\i}}a} \& {van
  Albada}}{{Gonz{\' a}lez-Garc{\'{\i}}a} \& {van
  Albada}}{2005}]{2005MNRAS.361.1043G}
{Gonz{\' a}lez-Garc{\'{\i}}a} A.~C.,  {van Albada} T.~S.,  2005, \mnras, 361,
  1043

\bibitem[\protect\citeauthoryear{{Graham}, {Lauer}, {Colless} \&
  {Postman}}{{Graham} et~al.}{1996}]{1996ApJ...465..534G}
{Graham} A.,  {Lauer} T.~R.,  {Colless} M.,    {Postman} M.,  1996, \apj, 465,
  534

\bibitem[\protect\citeauthoryear{{Graham}}{{Graham}}{2001}]{2001MNRAS.326..543%
G}
{Graham} A.~W.,  2001, \mnras, 326, 543

\bibitem[\protect\citeauthoryear{{Graham}, {Erwin}, {Caon} \&
  {Trujillo}}{{Graham} et~al.}{2001}]{2001ApJ...563L..11G}
{Graham} A.~W.,  {Erwin} P.,  {Caon} N.,    {Trujillo} I.,  2001, \apjl, 563,
  L11

\bibitem[\protect\citeauthoryear{{Graham}, {Trujillo} \& {Caon}}{{Graham}
  et~al.}{2001}]{2001AJ....122.1707G}
{Graham} A.~W.,  {Trujillo} I.,    {Caon} N.,  2001, \aj, 122, 1707

\bibitem[\protect\citeauthoryear{{Guti{\'e}rrez}, {Trujillo}, {Aguerri},
  {Graham} \& {Caon}}{{Guti{\'e}rrez} et~al.}{2004}]{2004ApJ...602..664G}
{Guti{\'e}rrez} C.~M.,  {Trujillo} I.,  {Aguerri} J.~A.~L.,  {Graham} A.~W.,
  {Caon} N.,  2004, \apj, 602, 664

\bibitem[\protect\citeauthoryear{{Hernquist}}{{Hernquist}}{1990}]{1990ApJ...35%
6..359H}
{Hernquist} L.,  1990, \apj, 356, 359

\bibitem[\protect\citeauthoryear{{Hernquist}}{{Hernquist}}{1992}]{1992ApJ...40%
0..460H}
{Hernquist} L.,  1992, \apj, 400, 460

\bibitem[\protect\citeauthoryear{{Hernquist}}{{Hernquist}}{1993a}]{1993ApJS...%
86..389H}
{Hernquist} L.,  1993a, \apjs, 86, 389

\bibitem[\protect\citeauthoryear{{Hernquist}}{{Hernquist}}{1993b}]{1993ApJ...4%
09..548H}
{Hernquist} L.,  1993b, \apj, 409, 548

\bibitem[\protect\citeauthoryear{{Heyl}, {Hernquist} \& {Spergel}}{{Heyl}
  et~al.}{1994}]{1994ApJ...427..165H}
{Heyl} J.~S.,  {Hernquist} L.,    {Spergel} D.~N.,  1994, \apj, 427, 165

\bibitem[\protect\citeauthoryear{{Jerjen} \& {Binggeli}}{{Jerjen} \&
  {Binggeli}}{1997}]{1997neg..conf..239J}
{Jerjen} H.,  {Binggeli} B.,  1997, in ASP Conf. Ser. 116: The Nature of
  Elliptical Galaxies; 2nd Stromlo Symposium {Are "Dwarf" Ellipticals Genuine
  Ellipticals?}.
pp 239--+

\bibitem[\protect\citeauthoryear{{Jesseit}, {Naab} \& {Burkert}}{{Jesseit}
  et~al.}{2005}]{2005MNRAS.360.1185J}
{Jesseit} R.,  {Naab} T.,    {Burkert} A.,  2005, \mnras, 360, 1185

\bibitem[\protect\citeauthoryear{{Jog} \& {Chitre}}{{Jog} \&
  {Chitre}}{2002}]{2002A&A...393L..89J}
{Jog} C.~J.,  {Chitre} A.,  2002, \aap, 393, L89

\bibitem[\protect\citeauthoryear{{Kawai}, {Fukushige}, {Makino} \&
  {Taiji}}{{Kawai} et~al.}{2000}]{2000PASJ...52..659K}
{Kawai} A.,  {Fukushige} T.,  {Makino} J.,    {Taiji} M.,  2000, \pasj, 52, 659

\bibitem[\protect\citeauthoryear{{Khochfar} \& {Burkert}}{{Khochfar} \&
  {Burkert}}{2006}]{2006A&A...445..403K}
{Khochfar} S.,  {Burkert} A.,  2006, \aap, 445, 403

\bibitem[\protect\citeauthoryear{{Labb{\' e}}, {Rudnick}, {Franx}, {Daddi},
  {van Dokkum}, {F{\" o}rster Schreiber}, {Kuijken}, {Moorwood}, {Rix}, {R{\"
  o}ttgering}, {Trujillo}, {van der Wel}, {van der Werf} \& {van
  Starkenburg}}{{Labb{\' e}} et~al.}{2003}]{2003ApJ...591L..95L}
{Labb{\' e}} I.,  {Rudnick} G.,  {Franx} M.,  {Daddi} E.,  {van Dokkum} P.~G.,
  {F{\" o}rster Schreiber} N.~M.,  {Kuijken} K.,  {Moorwood} A.,  {Rix} H.-W.,
  {R{\" o}ttgering} H.,  {Trujillo} I.,  {van der Wel} A.,  {van der Werf} P.,
    {van Starkenburg} L.,  2003, \apjl, 591, L95

\bibitem[\protect\citeauthoryear{{Merritt}, {Navarro}, {Ludlow} \&
  {Jenkins}}{{Merritt} et~al.}{2005}]{2005ApJ...624L..85M}
{Merritt} D.,  {Navarro} J.~F.,  {Ludlow} A.,    {Jenkins} A.,  2005, \apjl,
  624, L85

\bibitem[\protect\citeauthoryear{{Mihos} \& {Hernquist}}{{Mihos} \&
  {Hernquist}}{1996}]{1996ApJ...464..641M}
{Mihos} J.~C.,  {Hernquist} L.,  1996, \apj, 464, 641

\bibitem[\protect\citeauthoryear{{Naab} \& {Burkert}}{{Naab} \&
  {Burkert}}{2001a}]{2001ASPC..230..451N}
{Naab} T.,  {Burkert} A.,  2001a, in ASP Conf. Ser. 230: Galaxy Disks and Disk
  Galaxies {Gas Dynamics and Disk Formation in 3:1 Mergers}.
pp 451--452

\bibitem[\protect\citeauthoryear{{Naab} \& {Burkert}}{{Naab} \&
  {Burkert}}{2001b}]{2001ApJ...555L..91N}
{Naab} T.,  {Burkert} A.,  2001b, \apjl, 555, L91

\bibitem[\protect\citeauthoryear{{Naab} \& {Burkert}}{{Naab} \&
  {Burkert}}{2003}]{2003ApJ...597..893N}
{Naab} T.,  {Burkert} A.,  2003, \apj, 597, 893

\bibitem[\protect\citeauthoryear{{Naab}, {Burkert} \& {Hernquist}}{{Naab}
  et~al.}{1999}]{1999ApJ...523L.133N}
{Naab} T.,  {Burkert} A.,    {Hernquist} L.,  1999, \apjl, 523, L133

\bibitem[\protect\citeauthoryear{{Naab}, {Johansson}, {Efstathiou} \&
  {Ostriker}}{{Naab} et~al.}{2005}]{2005astro.ph.12235N}
{Naab} T.,  {Johansson} P.~H.,  {Efstathiou} G.,    {Ostriker} J.~P.,  2005,
  ArXiv Astrophysics e-prints

\bibitem[\protect\citeauthoryear{{Naab}, {Khochfar} \& {Burkert}}{{Naab}
  et~al.}{2006}]{2006ApJ...636L..81N}
{Naab} T.,  {Khochfar} S.,    {Burkert} A.,  2006, \apjl, 636, L81

\bibitem[\protect\citeauthoryear{{Naab} \& {Ostriker}}{{Naab} \&
  {Ostriker}}{2006}]{2006MNRAS.tmp...94N}
{Naab} T.,  {Ostriker} J.~P.,  2006, \mnras, pp 94--+

\bibitem[\protect\citeauthoryear{{Ostriker}}{{Ostriker}}{1980}]{1980ComAp...8.%
.177O}
{Ostriker} J.~P.,  1980, Comments on Astrophysics, 8, 177

\bibitem[\protect\citeauthoryear{{Peletier}, {Davies}, {Illingworth}, {Davis}
  \& {Cawson}}{{Peletier} et~al.}{1990}]{1990AJ....100.1091P}
{Peletier} R.~F.,  {Davies} R.~L.,  {Illingworth} G.~D.,  {Davis} L.~E.,
  {Cawson} M.,  1990, \aj, 100, 1091

\bibitem[\protect\citeauthoryear{{Pierce} \& {Tully}}{{Pierce} \&
  {Tully}}{1992}]{1992ApJ...387...47P}
{Pierce} M.~J.,  {Tully} R.~B.,  1992, \apj, 387, 47

\bibitem[\protect\citeauthoryear{{Press}, {Teukolsky}, {Vetterling} \&
  {Flannery}}{{Press} et~al.}{1992}]{1992nrfa.book.....P}
{Press} W.~H.,  {Teukolsky} S.~A.,  {Vetterling} W.~T.,    {Flannery} B.~P.,
  1992, {Numerical recipes in FORTRAN. The art of scientific computing}.
Cambridge: University Press, |c1992, 2nd ed.

\bibitem[\protect\citeauthoryear{{Prugniel} \& {Simien}}{{Prugniel} \&
  {Simien}}{1997}]{1997A&A...321..111P}
{Prugniel} P.,  {Simien} F.,  1997, \aap, 321, 111

\bibitem[\protect\citeauthoryear{{Rix}, {Carollo} \& {Freeman}}{{Rix}
  et~al.}{1999}]{1999ApJ...513L..25R}
{Rix} H.,  {Carollo} C.~M.,    {Freeman} K.,  1999, \apjl, 513, L25

\bibitem[\protect\citeauthoryear{{Rix}, {Franx}, {Fisher} \&
  {Illingworth}}{{Rix} et~al.}{1992}]{1992ApJ...400L...5R}
{Rix} H.,  {Franx} M.,  {Fisher} D.,    {Illingworth} G.,  1992, \apjl, 400, L5

\bibitem[\protect\citeauthoryear{{Robertson}, {Cox}, {Hernquist}, {Franx},
  {Hopkins}, {Martini} \& {Springel}}{{Robertson}
  et~al.}{2005}]{2005astro.ph.11053R}
{Robertson} B.,  {Cox} T.~J.,  {Hernquist} L.,  {Franx} M.,  {Hopkins} P.~F.,
  {Martini} P.,    {Springel} V.,  2005, ArXiv A e-prints

\bibitem[\protect\citeauthoryear{{Rothberg} \& {Joseph}}{{Rothberg} \&
  {Joseph}}{2004}]{2004AJ....128.2098R}
{Rothberg} B.,  {Joseph} R.~D.,  2004, \aj, 128, 2098

\bibitem[\protect\citeauthoryear{{Rubin}, {Graham} \& {Kenney}}{{Rubin}
  et~al.}{1992}]{1992ApJ...394L...9R}
{Rubin} V.~C.,  {Graham} J.~A.,    {Kenney} J.~D.~P.,  1992, \apjl, 394, L9

\bibitem[\protect\citeauthoryear{{Sersic}}{{Sersic}}{1968}]{1968adga.book.....%
S}
{Sersic} J.~L.,  1968, {Atlas de galaxias australes}.
Cordoba, Argentina: Observatorio Astronomico, 1968

\bibitem[\protect\citeauthoryear{{Springel}, {Di Matteo} \&
  {Hernquist}}{{Springel} et~al.}{2005}]{2005ApJ...620L..79S}
{Springel} V.,  {Di Matteo} T.,    {Hernquist} L.,  2005, \apjl, 620, L79

\bibitem[\protect\citeauthoryear{{Springel} \& {Hernquist}}{{Springel} \&
  {Hernquist}}{2005}]{2005ApJ...622L...9S}
{Springel} V.,  {Hernquist} L.,  2005, \apjl, 622, L9

\bibitem[\protect\citeauthoryear{{Trujillo}, {Aguerri}, {Guti{\' e}rrez} \&
  {Cepa}}{{Trujillo} et~al.}{2001}]{2001AJ....122...38T}
{Trujillo} I.,  {Aguerri} J.~A.~L.,  {Guti{\' e}rrez} C.~M.,    {Cepa} J.,
  2001, \aj, 122, 38

\bibitem[\protect\citeauthoryear{{Trujillo}, {Burkert} \& {Bell}}{{Trujillo}
  et~al.}{2004}]{2004ApJ...600L..39T}
{Trujillo} I.,  {Burkert} A.,    {Bell} E.~F.,  2004, \apjl, 600, L39

\bibitem[\protect\citeauthoryear{{Vazdekis}, {Trujillo} \& {Yamada}}{{Vazdekis}
  et~al.}{2004}]{2004ApJ...601L..33V}
{Vazdekis} A.,  {Trujillo} I.,    {Yamada} Y.,  2004, \apjl, 601, L33

\bibitem[\protect\citeauthoryear{{Weil} \& {Hernquist}}{{Weil} \&
  {Hernquist}}{1996}]{1996ApJ...460..101W}
{Weil} M.~L.,  {Hernquist} L.,  1996, \apj, 460, 101

\bibitem[\protect\citeauthoryear{{Young} \& {Currie}}{{Young} \&
  {Currie}}{1994}]{1994MNRAS.268L..11Y}
{Young} C.~K.,  {Currie} M.~J.,  1994, \mnras, 268, L11+

\bibitem[\protect\citeauthoryear{{Young} \& {Currie}}{{Young} \&
  {Currie}}{1995}]{1995MNRAS.273.1141Y}
{Young} C.~K.,  {Currie} M.~J.,  1995, \mnras, 273, 1141

\end{thebibliography}
\label{lastpage}

\end{document}